\documentclass[11pt]{article}

\usepackage[utf8]{inputenc}
\usepackage[T1]{fontenc}

\usepackage{graphicx}
\usepackage{float}
\usepackage{amsmath,amssymb,bm}
\usepackage[numbers]{natbib}
\usepackage[hidelinks]{hyperref}

\title{Topologically quantized macroscopic attractor states in hydrated DNA}
\author{
Mariusz Pietruszka\\
\small University of Silesia in Katowice, Faculty of Natural Sciences\\
\small Institute of Biology, Biotechnology and Environmental Protection\\
\small Katowice, Poland\\
\small Correspondence: \href{mailto:mariusz.pietruszka@us.edu.pl}{mariusz.pietruszka@us.edu.pl}
}

\date{\today}

\begin{document}

\maketitle

\begin{abstract}

Driven dissipative systems at ambient conditions typically exhibit continuous responses shaped by fluctuations and relaxation, with discrete macroscopic states arising only under specific dynamical constraints. Here, we report the emergence of discrete attractor states in a quasi-two-dimensional hydrated DNA sample under magnetic excitation. The transverse polarization voltage $V_{xy}$ displays telegraph switching between well-defined levels, indicating stochastic transitions between metastable macroscopic states. Statistical analysis of the voltage time series reveals bimodal distributions and strong Bayesian model selection in favor of multiple coexisting states. A phenomenological phase-field framework describes the observed metastability and switching in terms of a proposed collective $U(1)$ polarization phase. Within this interpretation, phase-slip dynamics and integer-labeled winding sectors provide a possible organizing principle for the discrete
macroscopic response. The results suggest that macroscopic quantization can emerge in a classical system at ambient conditions as a consequence of dissipative dynamics constrained by phase topology.

\end{abstract} % Sci Rep

\section{Introduction}

Classical driven dissipative systems at ambient conditions typically exhibit continuous responses governed by fluctuations, relaxation, and nonlinear feedback. Under certain dynamical constraints, however, such systems can develop multiple metastable states and complex attractor structures. An important open question is whether soft-matter systems can support discrete macroscopic states that are robust under realistic environmental conditions. Here, we report that a hydrated DNA system subjected to magnetic excitation exhibits telegraph switching between discrete levels of transverse polarization voltage $V_{xy}$, indicating stochastic transitions between metastable macroscopic attractors. Statistical analysis of the voltage time series supports the presence of distinct states with well-defined distributions. We interpret these observations within a topological phase-field framework in which a collective $U(1)$ polarization phase organizes into integer-labeled winding sectors, with transitions mediated by phase-slip events. These results demonstrate that discrete macroscopic states can emerge in a classical system as a consequence of dissipative dynamics constrained by phase topology, providing a route to topological phase behavior in soft matter under ambient conditions.

DNA is not only a genetic information carrier but also a structurally and dynamically complex polyelectrolyte whose physical properties are strongly influenced by hydration and ionic environment. The double-helical backbone supports extended hydrogen-bond networks and collective vibrational modes that can couple to surrounding water molecules and mobile protons. Theoretical descriptions of DNA stability and dynamics
\cite{Prohofsky1995}, broader discussions of coherence in
biological matter \cite{Pokorny2014}, and studies of
DNA-mediated charge transport \cite{Barton2011} provide
complementary contexts for investigating its electrical response. In aqueous environments the DNA–water complex therefore forms a mesoscale soft-matter system capable of supporting coupled electromechanical and polarization excitations. These properties make hydrated DNA a natural platform for exploring driven collective phenomena and emergent dynamical states under external fields. In thin hydrated layers confined between a substrate and a cover glass, the DNA–water complex effectively forms a quasi-two-dimensional soft-matter system in which collective polarization and transport processes are constrained to a planar geometry, making it particularly suitable for studying driven phase dynamics under external fields. The persistence of this behavior in the presence of 0.9\% NaCl demonstrates that the observed bistability is not restricted to idealized hydration conditions, but remains robust under ionic screening environments closer to biological systems \cite{Pietruszka2026BistableDNA}. Moreover, telegraph-like switching under moderate magnetic-field excitation and over a broad temperature range was already documented in our previously published study on hydrated DNA, where the transverse signal alternated between two preferred macroscopic levels, indicating bistable collective polarization dynamics \cite{Pietruszka2026BioSystems}.

Classical driven and dissipative systems can generate complex dynamical structures, including coherent collective modes, synchronization phenomena, and long-lived metastable states. In soft condensed matter and biological materials, such behaviour is often associated with nonlinear interactions and energy dissipation rather than microscopic quantum coherence. Early theoretical work proposed that biomolecular assemblies may support collective vibrational modes under non-equilibrium conditions, a mechanism often discussed in the context of Fr\"ohlich-type coherent excitations \cite{Frohlich1968,froehlich1980}. More generally, dissipative field theories predict the emergence of ordered macroscopic states and attractor dynamics in systems coupled to an environment \cite{anderson1995,Vitiello2001}. In such frameworks, discrete macroscopic states can arise not from microscopic energy quantization, but from the topology of the underlying phase field, where phase-slip transitions separate integer-labelled sectors \cite{Kosterlitz1973}. However, identifying experimental realizations of such topologically stabilized dynamical states in biological soft matter remains an open challenge. It is widely recognized that microscopic quantum coherence in biological systems is expected to decohere extremely rapidly under ambient conditions \cite{Tegmark2000}, although possible roles of quantum effects in biological processes continue to be actively explored \cite{McFadden2014}. The phenomenon studied here does not rely on long-lived microscopic quantum states, but instead involves collective polarization dynamics in a driven, dissipative medium, where discrete macroscopic states emerge from the topology of a compact phase field. 

Discrete macroscopic states in driven classical systems are known in several well-established contexts, including phase-locked Josephson junction arrays exhibiting voltage steps under external driving \cite{Fazio2001}, synchronization phenomena in nonlinear oscillators and injection-locked resonators \cite{Pikovsky2003}, pattern selection in parametrically driven fluids such as Faraday waves \cite{Cross1993}, and random telegraph switching in mesoscopic electronic systems associated with bistable defects or traps \cite{Dutta1981RMP}. In most of these cases, however, the observed discreteness arises under specialized conditions, such as cryogenic temperatures (as in superconducting Josephson devices), engineered resonant structures or oscillators, electronic circuits with designed nonlinear elements, or externally imposed synchronization protocols. 

The system investigated here differs in several important respects. The discrete states emerge in a hydrated biological polymer operating at ambient temperature, without engineered resonators or externally imposed phase locking. Instead, the relevant dynamical variable is a collective polarization mode supported by the hydrated DNA matrix and its surrounding hydrogen-bond network. The resulting behavior, therefore, represents an example of discrete macroscopic states arising in a soft biological medium through collective phase dynamics under ordinary environmental conditions. This combination of ambient operation, biological soft matter, and collective polarization dynamics makes the phenomenon unusual among classical driven systems.

In this work, we investigate hydrated DNA subjected to magnetic excitation under ambient laboratory conditions. The transverse polarization voltage $V_{xy}$ exhibits intermittent telegraph switching between well-defined levels, indicating stochastic transitions between discrete metastable states of the system. Statistical analysis of the voltage time series reveals bimodal voltage distributions and a decisive model-selection favoring multiple attractors. These observations can be interpreted using a minimal topological phase-field description in which a collective $U(1)$ polarization phase forms integer-labeled winding sectors connected by phase-slip events, with the observed voltage levels corresponding to macroscopic attractor states of a driven dissipative field.

Together, these observations point to a classical driven–dissipative system in which telegraph switching arises from competition between metastable attractor states belonging to distinct topological sectors of a collective $U(1)$ phase field, with transitions mediated by phase slips $\Delta\phi = \pm 2\pi$, implying macroscopic phase quantization of topological origin rather than discrete energetic quantization. Hence, in this study, the term ``quantized'' is used operationally to refer to reproducible, discrete voltage levels separated by well-defined step intervals, without implying the existence of discrete microscopic energy levels.

\section{Results}

We report discrete polarization states and telegraph-like switching in a driven soft-matter system realized in a quasi-two-dimensional DNA--water sample under magnetic-field excitation at ambient conditions.
Above a well-defined threshold, the transverse voltage organizes into a sequence of robust plateaus separated by fluctuation-dominated transition regions.
Statistical analysis reveals long-lived metastable states with unimodal voltage distributions on plateaus and bimodal distributions near step boundaries; Bayesian model selection in representative transition windows provides strong statistical support (e.g., $\Delta\mathrm{BIC}\sim10^{2}$) for bimodality, consistent with real-time competition between neighboring attractor states.
The persistence of discrete states at room temperature is difficult to reconcile with interpretations based solely on microscopic energy quantization.
Instead, the observations are consistent with overdamped phase dynamics of a driven--dissipative collective polarization field, in which a compact phase variable supports metastable attractors and noise-assisted phase-slip transitions.

Here, a plateau denotes an approximately constant voltage
level over a finite interval of the field sweep.
A metastable state denotes a preferred response level
occupied for a finite time before a transition.
An attractor refers to its proposed dynamical representation
in the phenomenological model, while a polarization state
specifies the proposed physical interpretation.
Plateau widths in magnetic-field units are distinct from
residence times measured directly from the time traces.

\subsection{Magnetic-field threshold and discrete voltage levels}

We first characterize the global magnetic-field response before examining the fine structure of the post-threshold regime and its statistical properties.

\begin{figure}[H]
    \centering
    \includegraphics[width=0.85\linewidth]{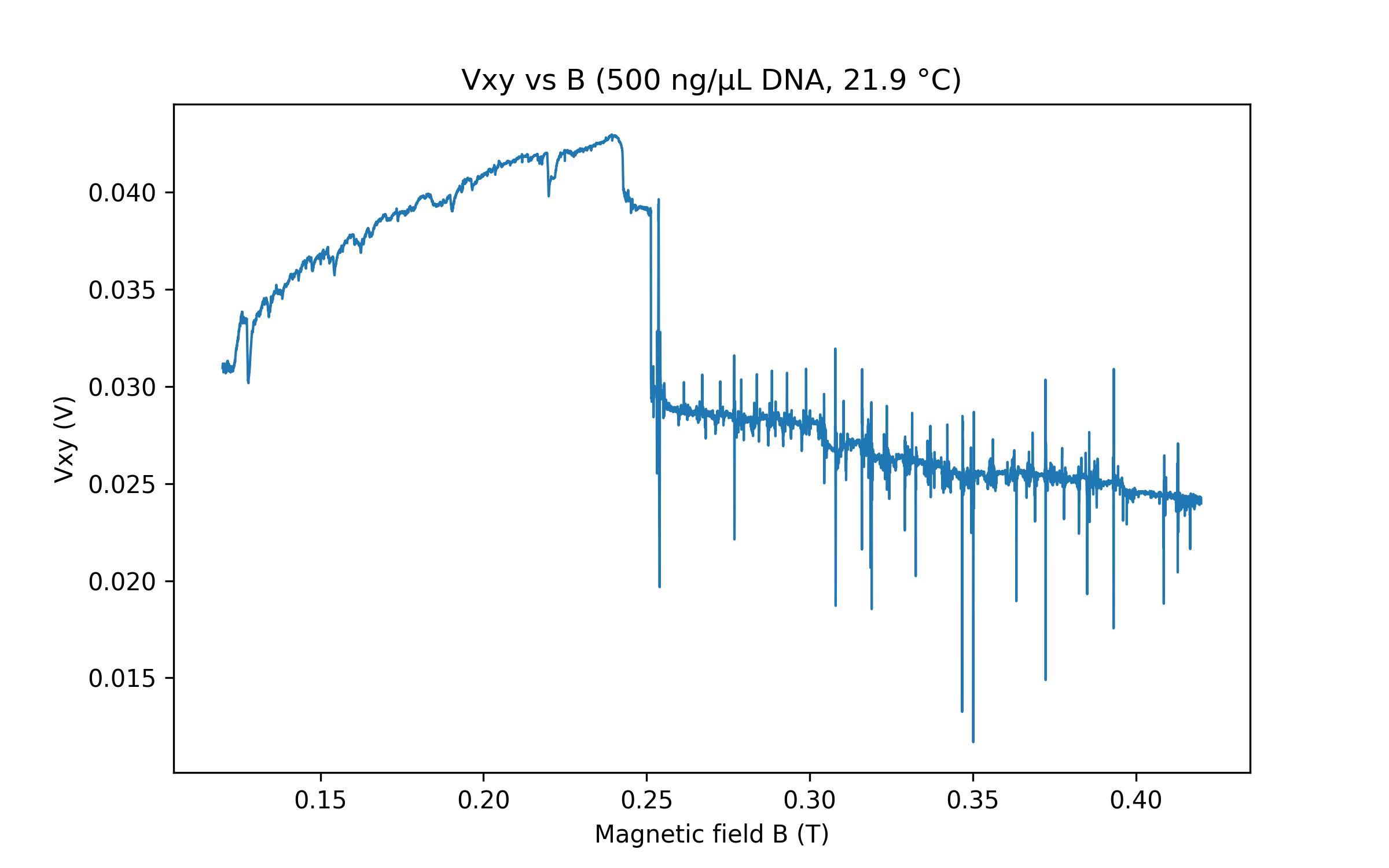}
    \caption{
Transverse voltage $V_{xy}$ measured during a magnetic-field sweep in hydrated DNA (500 ng µL$^{-1}$, 5 µL) at ambient temperature ($T=21.9^{\circ}$C). The underlying time record contains 20\,398 voltage samples
acquired over approximately 700~s.
The magnetic field was adjusted manually and increased monotonically during the measurement.
The plateaus are not synchronized with individual field adjustments but correspond to intrinsic metastable attractors stabilized under sufficiently quiet magnetic conditions.
Small cyclic fluctuations accompany field variation, while strong telegraph-switching events occur rarely and are confined to plateau boundaries.
A pronounced threshold appears near $B \approx 0.25$--0.27~T, marking a transition from a weakly varying response to a regime characterized by strong nonlinearity and the emergence of discrete voltage structures.
The full dataset is available on Zenodo 10.5281/zenodo.14716955. This figure presents a new visualization of the experimental
record previously reported in Ref.~\cite{Pietruszka2025PhysicaB}.
The present work examines its metastable-state organization
and switching statistics.
}
    \label{fig:threshold}
\end{figure}

\noindent
Fig.~\ref{fig:threshold} presents the transverse voltage $V_{xy}$ measured during a monotonic magnetic-field sweep at ambient temperature ($T=21.9^{\circ}$C).
A pronounced threshold occurs near $B \approx 0.25$--0.27~T, indicating the onset of a qualitatively different response regime.
Above this threshold, $V_{xy}$ no longer evolves smoothly but develops discrete structures and strong intermittent fluctuations.

\begin{figure}[H]
    \centering
    \includegraphics[width=0.75\linewidth]{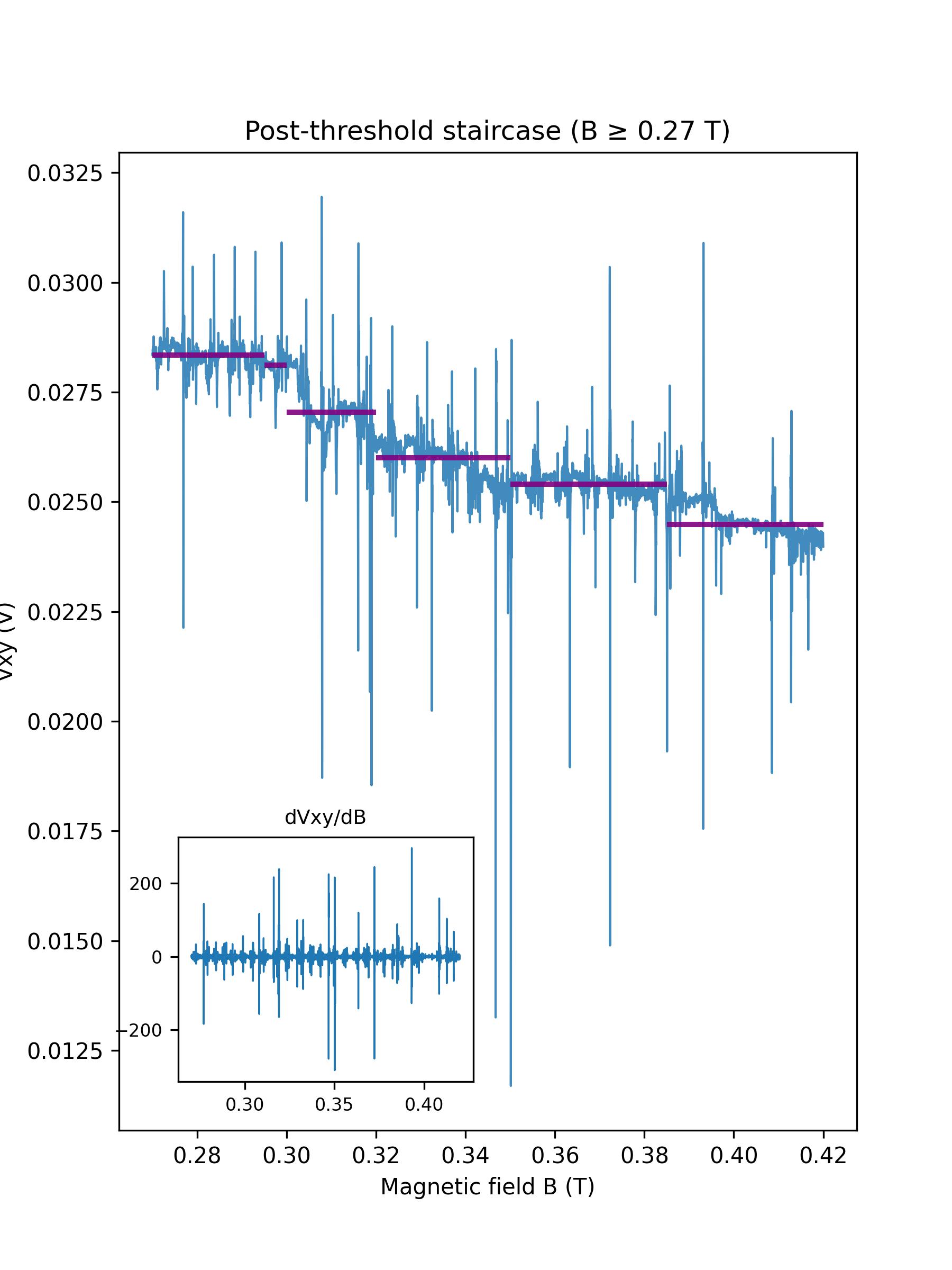}
    \caption{
Zoomed view of the post-threshold transverse voltage response $V_{xy}(B)$ from Fig.~\ref{fig:threshold}, focusing on the magnetic-field range $B \gtrsim 0.27$~T.
In this regime, the signal organizes into a sequence of discrete, metastable voltage plateaus separated by fluctuation-dominated transition regions.
Horizontal line segments indicate representative plateau levels, determined from the local $V_{xy}$ values between successive extrema in the field derivative.
The finite magnetic-field extent of each segment reflects the limited stability range of the corresponding metastable state.
\textit{Inset:} field derivative $dV_{xy}/dB$, highlighting sharp transition events and intermittent telegraph-like switching between neighboring plateaus.
}
    \label{fig:staircase}
\end{figure}

\noindent
A zoomed view of the post-threshold regime (Fig.~\ref{fig:staircase}) reveals a staircase-like organization into metastable plateaus separated by sharp transition regions.
The derivative $dV_{xy}/dB$ (inset) highlights localized switching events, while the plateau intervals are comparatively quiescent.
Near step boundaries, the signal frequently exhibits telegraph-like switching between neighboring levels, consistent with competition between coexisting metastable states and noise-assisted transitions.

\begin{figure}[H]
    \centering
    \includegraphics[width=0.75\linewidth]{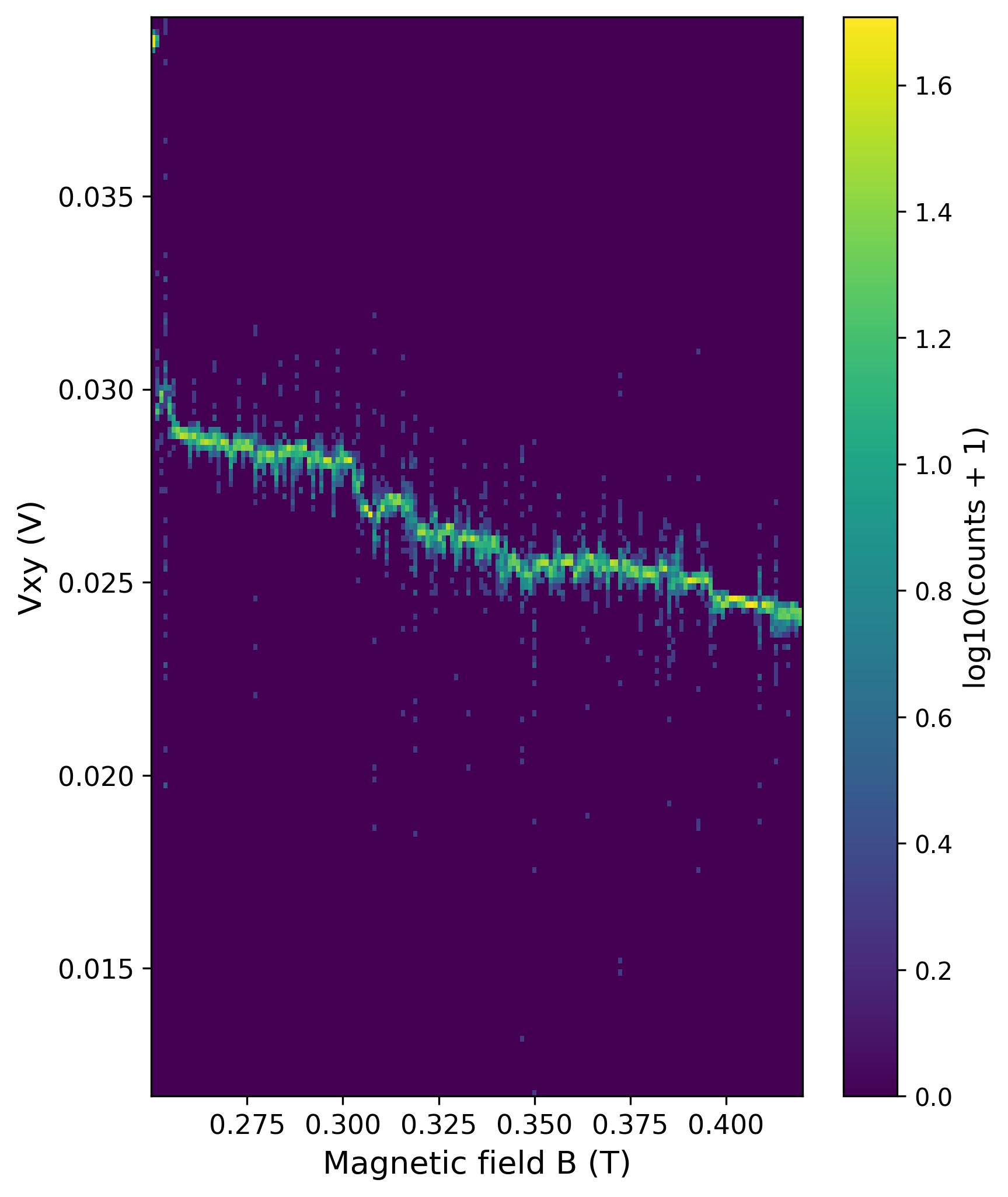}
    \caption{
Dwell-density map of the transverse voltage $V_{xy}$ as a function of magnetic field $B$ for the DNA--water system (500~ng~$\mu$L$^{-1}$, $T=21.9^{\circ}$C), constructed as a two-dimensional histogram of the time-ordered data in the post-threshold regime ($B \gtrsim 0.25$~T).
Color encodes $\log_{10}(\mathrm{counts}+1)$ and highlights the most frequently occupied voltage levels.
The appearance of narrow, ridge-like bands indicates a discrete set of preferred metastable polarization states persisting over finite magnetic-field intervals.
Broadening and fragmentation of the ridges near their boundaries reflect enhanced fluctuations and telegraph-like switching associated with transitions between neighboring states.
}
    \label{fig:density}
\end{figure}

\noindent
To quantify the discreteness of the response independently of any single trajectory, Fig.~\ref{fig:density} shows the dwell-density distribution of $V_{xy}$ as a function of $B$ constructed from the time-ordered data.
The appearance of narrow ridge-like bands demonstrates that the system repeatedly occupies a discrete set of preferred voltage states over finite field intervals, with enhanced broadening at ridge boundaries where switching activity is most pronounced.
The persistence of these discrete states at ambient temperature supports an interpretation in terms of topological phase quantization of a collective polarization field, rather than quantized energy levels.

\begin{figure}[H]
    \centering
    \includegraphics[width=1\linewidth]{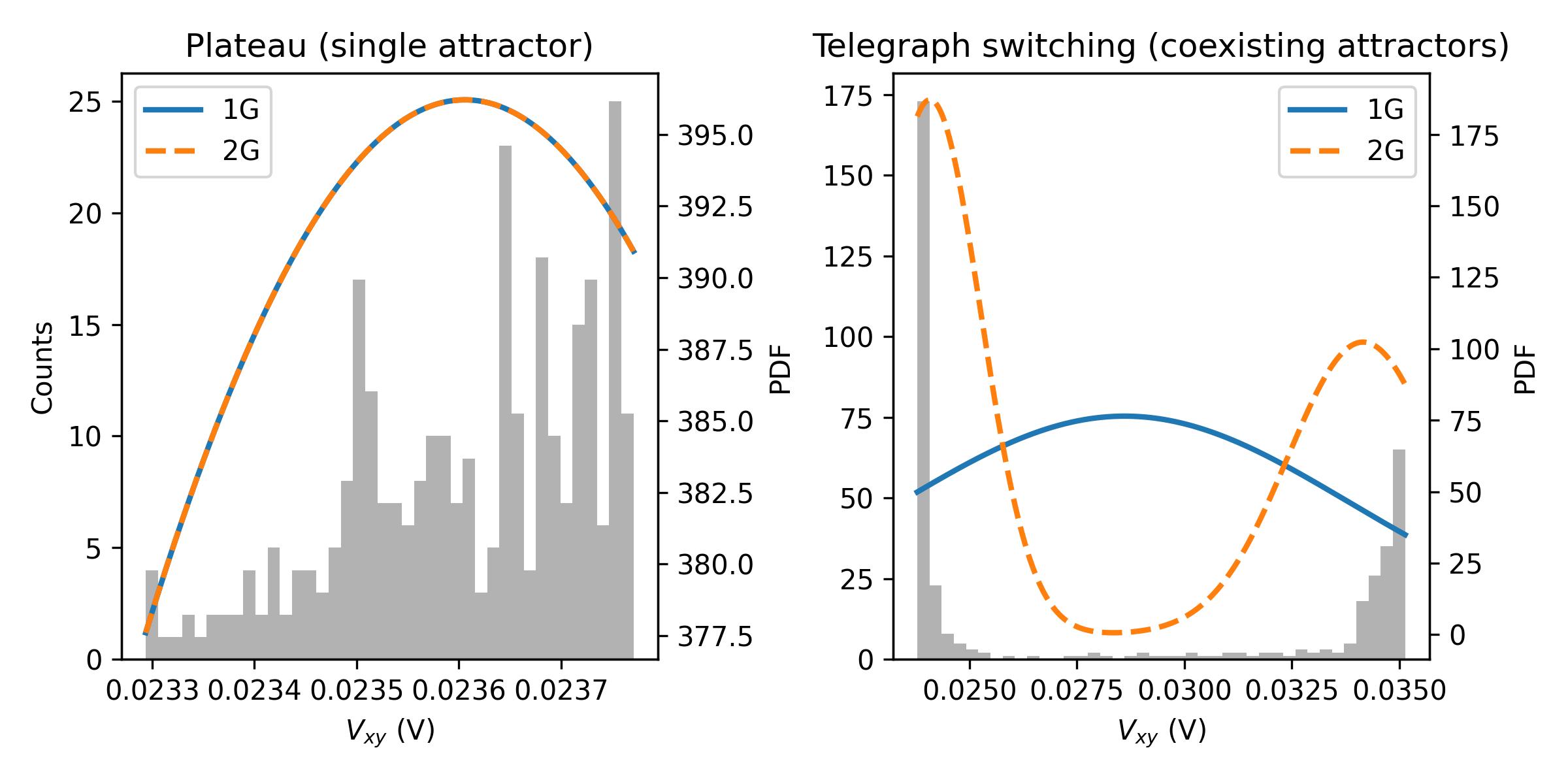}
    \caption{
Statistical signature of attractor competition in the transverse voltage response.
Left: within a stable plateau, the voltage distribution is unimodal and well described
by a single Gaussian probability density, indicating residence near a single
metastable attractor.
Right: near a plateau boundary, the distribution becomes bimodal
and is described by a two-Gaussian mixture, consistent with
time-resolved switching between two preferred voltage levels. Model selection using the Bayesian information criterion confirms a transition from unimodal to bimodal voltage distributions near plateau boundaries.
Field-resolved Bayesian model selection is presented
in Fig.~\ref{fig:bic_overlay}.
}
    \label{fig:attractors}
\end{figure}

Building on our earlier characterization of hydrated DNA
\cite{Pietruszka2026BioSystems}, the present measurements
resolve switching between preferred transverse-voltage levels.
Near plateau boundaries, Bayesian model comparisons yield
$\Delta\mathrm{BIC}\sim10^{2}$ in favor of bimodal distributions.
Together with the time traces, these statistics support
local switching between metastable response states.
Within the proposed collective-polarization model, this behavior
can be represented as transitions between neighboring attractor
basins in a driven dissipative landscape.

% Wstawione po wysłaniu do Sci Rep / użyte też w Science Advances - tam usunąć przed publikacją

\begin{figure}[H]
\centering
\includegraphics[width=.8\linewidth]{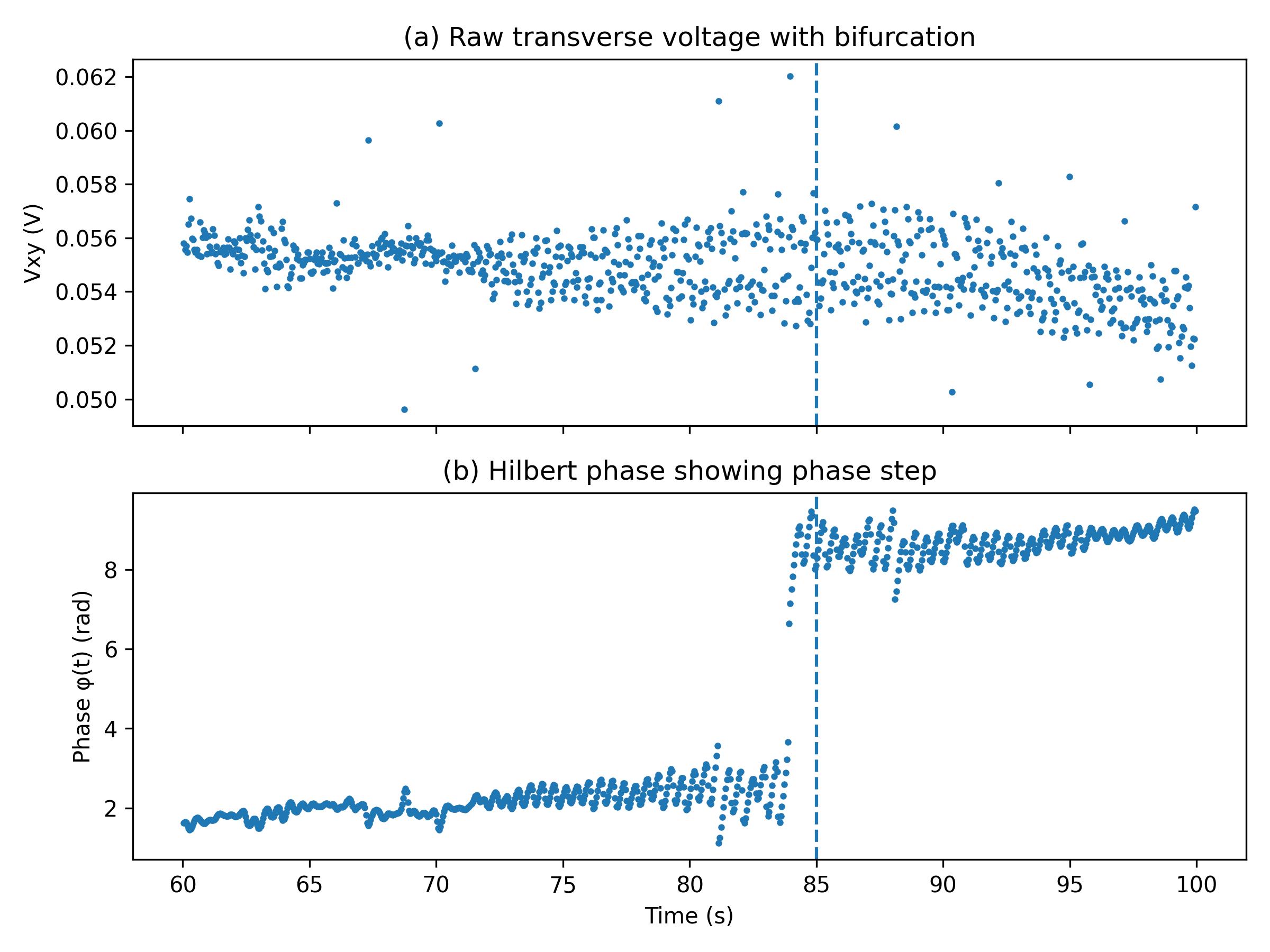}
\caption{
\textbf{Transient reconfiguration event used for the Hilbert-phase analysis.}
(a) Raw transverse-voltage signal $V_{xy}(t)$ showing bifurcation into two metastable branches near $t\approx70$ s.
(b) Unwrapped Hilbert phase reconstructed from the same signal. A delayed phase reconfiguration occurs near $t\approx85$ s, approximately 15 s after branch formation. Enhanced fluctuations preceding the transition are consistent with critical slowing down, while reduced fluctuations afterward indicate relaxation into a new metastable phase sector. The observed phase jump is consistent with a phase-slip event of order $\Delta\phi \approx 2\pi$, linking the transient dynamics to transitions between neighboring topological phase sectors. The vertical dashed line marks the phase-transition time.
}
\label{fig:appendix_raw_bifurcation}
\end{figure}

Figure~\ref{fig:appendix_raw_bifurcation} presents the transient
reconfiguration event during a magnetic-field ramp from
$B=0.35$ to $0.45$~T at $T\approx13.8^\circ$C, showing the raw
transverse voltage and the unwrapped Hilbert phase reconstructed
from the same signal. The voltage trace
develops two branches near $t\approx70$~s, followed by a phase
reconfiguration near $t\approx85$~s. The approximately 15~s delay
distinguishes the onset of voltage branching from the subsequent
change in the reconstructed phase. The phase increment is approximately
$2\pi$, consistent with a phase-slip-like event within the proposed
collective phase description.

\subsection{Bayesian analysis of voltage bistability}

\begin{figure}[H]
\centering
\includegraphics[width=\linewidth]{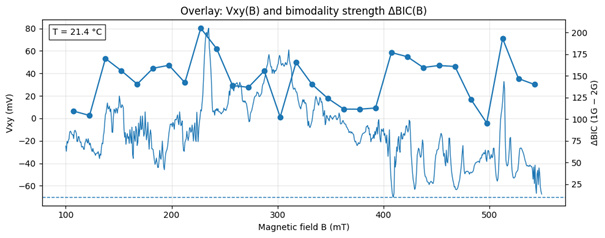}
\caption{
Overlay of the transverse voltage $V_{xy}(B)$ (left axis) and the bimodality strength $\Delta\mathrm{BIC}(B)$ (right axis) obtained from Bayesian model selection. The transverse voltage is shown as a continuous curve, while circular markers indicate $\Delta\mathrm{BIC}$ computed from Gaussian mixture fits within 15~mT field windows. Positive values $\Delta\mathrm{BIC}>10$ provide strong statistical evidence for bimodality; pronounced maxima occur at quasi--plateau boundaries, indicating enhanced attractor competition and intermittent switching in the mixed ohmic--polarization regime.
}
\label{fig:bic_overlay}
\end{figure}

To quantitatively assess the presence of competing macroscopic states underlying the observed telegraph switching, we performed a Bayesian information--criterion (BIC) analysis of transverse voltage histograms acquired during a linear magnetic--field sweep.

The transverse polarization voltage $V_{xy}(B)$ was measured in a hydrated DNA--water sample (500~ng~$\mu$L$^{-1}$, 5~$\mu$L volume, $0.7\times0.7$~cm$^{2}$ contact geometry) at $T=21.4\,^{\circ}$C under ambient laboratory conditions. The magnetic field was swept continuously from 100 to 550~mT. For each magnetic--field interval of width 15~mT, a voltage histogram was constructed and fitted using both a single--Gaussian (1G) model and a two--Gaussian mixture (2G) model. Model selection was performed using the Bayesian information criterion,
\begin{equation}
\Delta\mathrm{BIC} = \mathrm{BIC}_{1\mathrm{G}} - \mathrm{BIC}_{2\mathrm{G}},
\end{equation}
where positive values favor the bimodal model. Following standard criteria, $\Delta\mathrm{BIC}>10$ provides strong statistical evidence for bimodality.

Figure~\ref{fig:bic_overlay} displays the raw transverse voltage $V_{xy}(B)$ together with the corresponding $\Delta\mathrm{BIC}(B)$. Pronounced maxima in $\Delta\mathrm{BIC}$, reaching values well above 100 at specific magnetic fields, occur systematically at the boundaries between quasi--plateau regions of $V_{xy}$. These peaks indicate coexistence of two distinct transverse-polarization states and quantitatively demonstrate competition between metastable attractors in the mixed ohmic--polarization regime.

Independent time--domain analysis of $V_{xy}$ near these field values reveals telegraph--like switching between two discrete voltage levels, with characteristic dwell times of several seconds. Notably, the locations of enhanced $\Delta\mathrm{BIC}$ coincide with intervals exhibiting telegraph switching, confirming that histogram bimodality arises from real--time alternation between competing macroscopic polarization states rather than measurement noise. The observed dynamics are consistent with noise--assisted transitions between neighboring metastable attractors, reflecting intermittent basin competition prior to stabilization of a single polarization state.

The large voltage separation between the two telegraph states ($\sim 20$~mV), combined with much smaller intra-plateau fluctuations, yields decisively bimodal voltage distributions ($\Delta\mathrm{BIC} \gg 100$). Such behavior is characteristic of switching between metastable attractors in driven dissipative systems. Within the phase-field framework adopted here, this phenomenology is consistent with discrete, integer-labeled polarization states. It may be viewed as suggestive of phase attractors associated with distinct topological sectors of the collective polarization field.

To make the statistical evidence more transparent, we compare representative voltage histograms from two field intervals: a stable plateau window and a quasi--plateau boundary window. In the plateau window the voltage distribution is unimodal and well described by a single Gaussian, consistent with residence in a single attractor basin. In contrast, the boundary window exhibits a clearly bimodal distribution captured by a two--Gaussian mixture model with large positive $\Delta\mathrm{BIC}$. The appearance of bimodality only within the switching interval confirms that the observed voltage statistics arise from stochastic transitions between competing metastable attractors rather than from broadened single-state fluctuations.

A separate stationary-field experiment at $B=0.35$~T
and $T=13.1\pm0.1^\circ$C resolves switching between
metastable Hilbert-phase states reconstructed from $V_{xy}(t)$
\cite{Pietruszka2026PhaseSwitching}.
Gaussian-mixture classification of the switching interval
yields a mean residence time of 9.19~s, with
$\Delta\mathrm{BIC}=1068$ favoring the two-component phase
distribution. This complementary result demonstrates
that switching also occurs without magnetic-field sweeping.

\subsection{Complementary voltage and photovoltaic measurements}

\noindent\textbf{Experimental conditions:} Hydrated genomic DNA (500~ng~$\mu$L$^{-1}$, 5~$\mu$L), 0.7$\times$0.7~cm$^{2}$ substrate, $T=20.3^{\circ}$C, RH $\approx26\%$, dark, continuous linear magnetic-field sweep.

\subsubsection{Time-domain signatures of telegraph switching}

The transverse voltage $V_{xy}(t)$ exhibits pronounced non-Gaussian fluctuations characterized by abrupt transitions between discrete voltage levels. These transitions occur irregularly in time, with dwell periods ranging from several seconds to over one minute. Such behavior is characteristic of telegraph switching between metastable states.

Importantly, no comparable switching is observed in the longitudinal voltage or in the current-monitoring shunt channel, confirming that the effect is intrinsic to the transverse response rather than arising from instrumental or current instabilities.

\subsubsection{Spectral analysis of transverse and photovoltaic signals}

To characterize the temporal structure of the switching dynamics, Fast Fourier Transform (FFT) analysis was applied to both the transverse voltage $V_{xy}(t)$ and the simultaneously recorded photovoltaic (PV) signal.

The discrete Fourier transform is defined as
\begin{equation}
\tilde{V}(f) = \sum_{n=0}^{N-1} V(t_n)\, e^{-2\pi i f t_n},
\end{equation}
where $t_n$ denotes uniformly sampled time points and $f$ is the frequency.

\begin{figure}[H]
\centering
\includegraphics[width=0.95\linewidth]{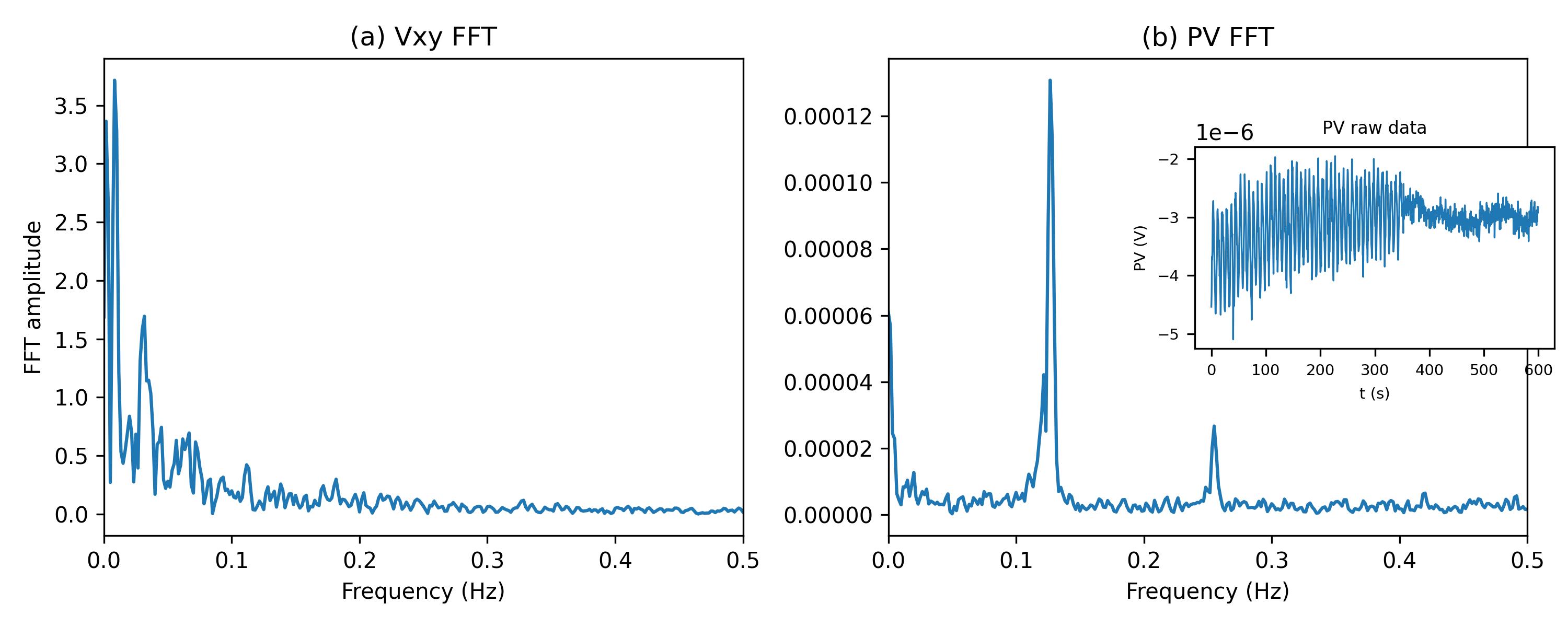}
\caption{Frequency-domain analysis of transverse voltage and photovoltaic signals at $T=20.3^\circ$C. 
(a) FFT amplitude spectrum of $V_{xy}$ after detrending and windowing, dominated by low-frequency components associated with slow collective dynamics. 
(b) FFT spectrum of the photovoltaic (PV) signal, showing a narrow peak near $f\approx0.13$~Hz. Inset: raw PV time trace acquired simultaneously during the magnetic-field sweep.}
\label{fig:S2}
\end{figure}

As shown in Fig.~\ref{fig:S2}, the $V_{xy}$ spectrum is dominated by low-frequency components below approximately 0.1--0.2 Hz, consistent with slow stochastic switching between metastable states rather than harmonic oscillations. In contrast, the PV signal displays a well-defined spectral peak near $f \approx 0.13$ Hz, corresponding to a periodic modulation with a characteristic timescale of $\sim$8 s. The absence of strict phase locking between the two signals indicates that the transverse switching dynamics are not trivially driven by the PV oscillation.

\subsubsection{Statistical evidence for discrete transverse states}

To quantify the apparent discreteness of the transverse response, probability distributions of $V_{xy}$ were constructed over the full magnetic-field sweep. The resulting histogram deviates strongly from a single Gaussian distribution and is instead well described by a two-component Gaussian mixture model,
\begin{equation}
P(V_{xy}) = \sum_{i=1}^{2} w_i\, \mathcal{N}(\mu_i, \sigma_i^2),
\end{equation}
where $w_i$ are statistical weights and $\mu_i$ the mean voltages of the two states.

\begin{figure}[H]
\centering
\includegraphics[width=0.9\linewidth]{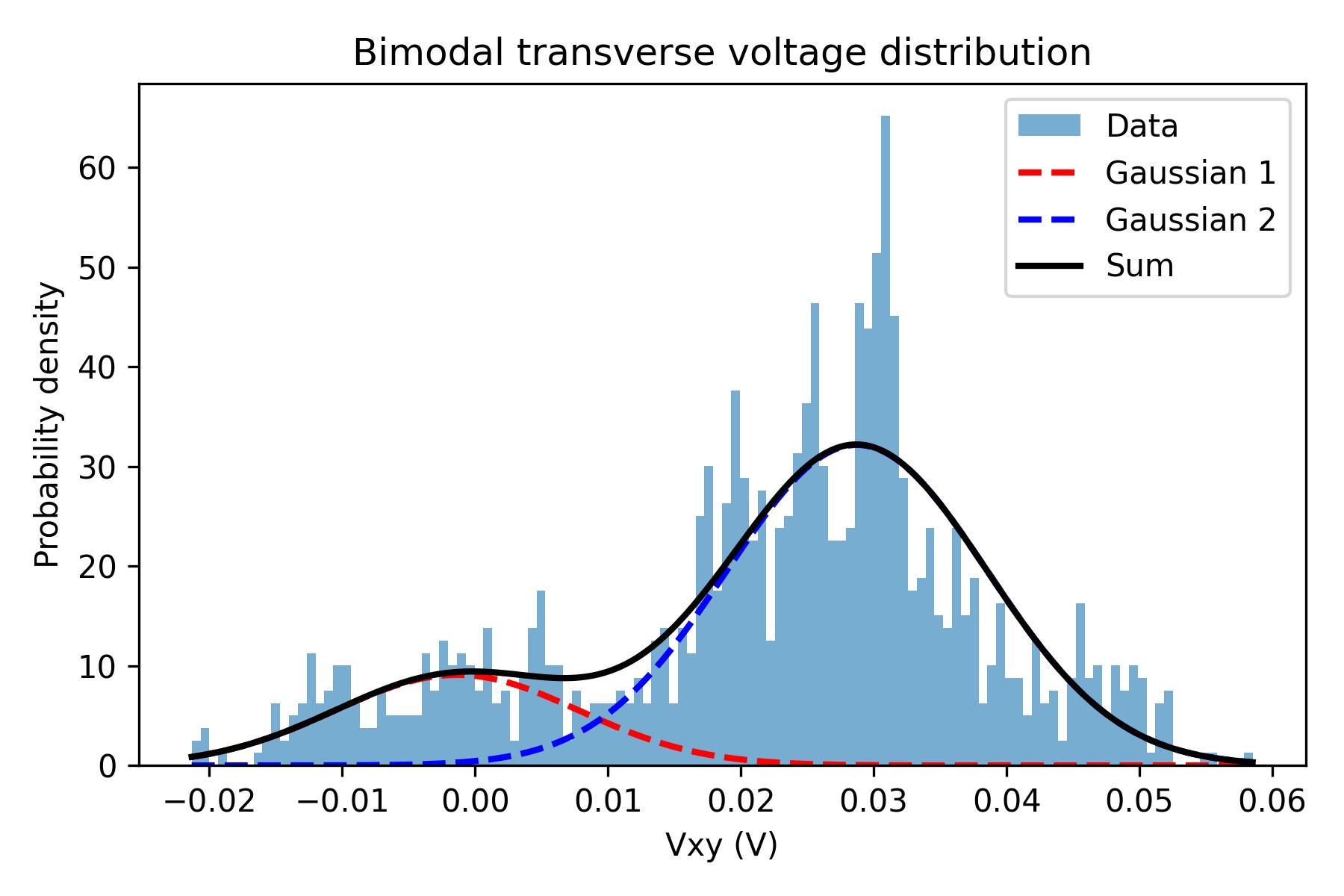}
\caption{Histogram of the transverse voltage $V_{xy}$ over the full magnetic-field sweep. The bimodal distribution demonstrates the existence of two discrete transverse voltage states. Dashed curves show the Gaussian components of a two-state mixture fit; the solid curve shows their sum.}
\label{fig:S3}
\end{figure}

A two-Gaussian mixture fit yields state means $\mu_1\approx -1.37~\mathrm{mV}$ and $\mu_2\approx +28.73~\mathrm{mV}$ (separation $\Delta\mu\approx 30~\mathrm{mV}$), with unequal statistical weights $w_1\approx 0.21$ and $w_2\approx 0.79$, consistent with different dwell times in the two states. These values remain essentially independent of magnetic field, while the occupation probability of each state varies strongly with $B$. The resulting bimodal histogram and Gaussian-mixture fit are shown in Fig.~\ref{fig:S3}.

\subsubsection{Time-resolved attractor dynamics}
To examine the temporal evolution of the two states, the data were analyzed using sliding time windows. Within each window, a two-state Gaussian mixture model was fitted, allowing the instantaneous positions of the low- and high-voltage states to be tracked.

\begin{figure}[H]
\centering
\includegraphics[width=0.9\linewidth]{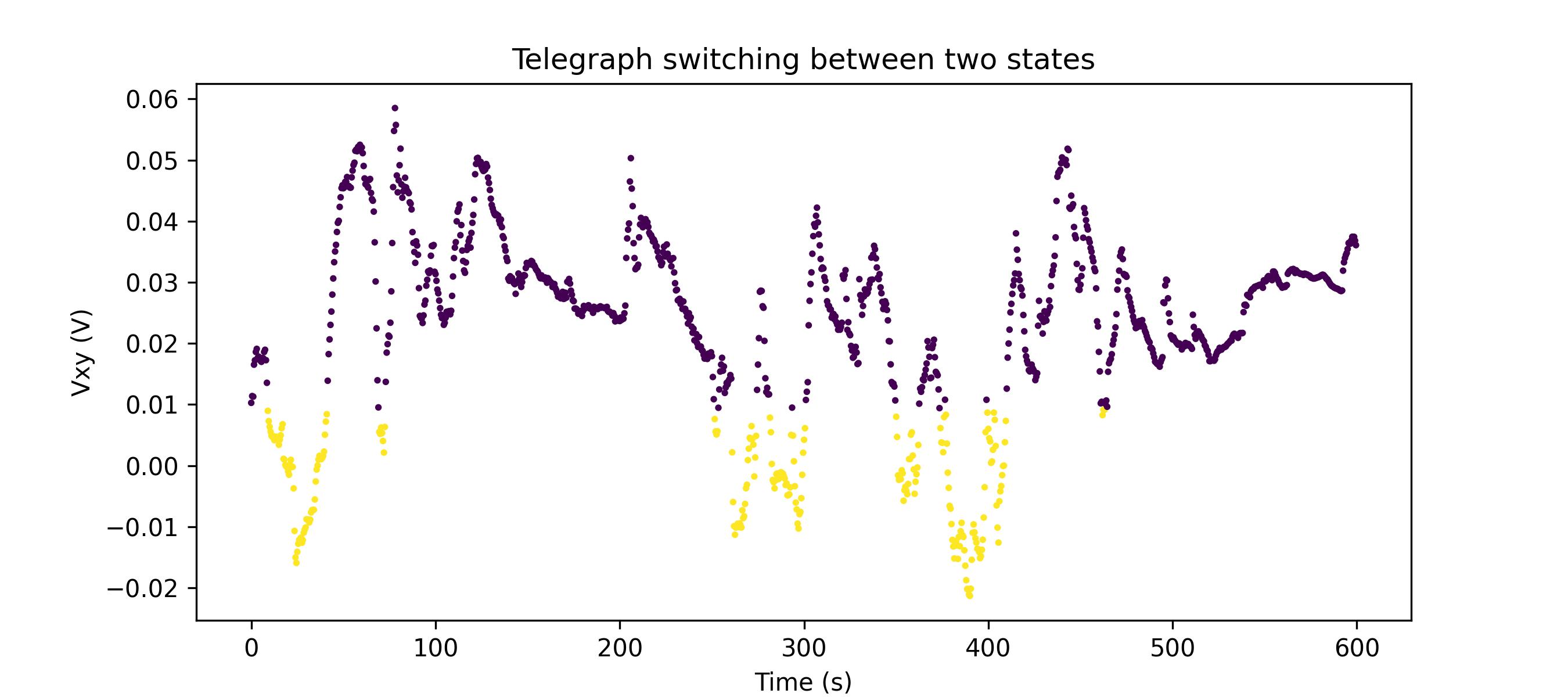}
\caption{Sliding-window tracking of metastable transverse voltage states. The two branches correspond to coexisting attractors; intermittent crossings indicate stochastic telegraph switching events.}
\label{fig:S4}
\end{figure}

The sliding-window tracking of the two metastable branches is shown in Fig.~\ref{fig:S4}.

This analysis reveals that the voltage values of the two states remain approximately constant throughout the sweep, while transitions between them occur intermittently. The switching probability increases in intermediate magnetic-field regions and is suppressed at higher fields, where one state becomes dominant.

\subsubsection{Transition rates and telegraph statistics}

The telegraph process may be modeled as a two-state Markov system characterized by transition rates $k_{01}$ and $k_{10}$ between the low (0) and high (1) states. From the classified time series, transition rates were extracted according to
\begin{equation}
k_{ij} = \frac{N_{i\rightarrow j}}{N_i\, \Delta t},
\end{equation}
where $N_{i\rightarrow j}$ is the number of observed transitions, $N_i$ the number of samples in state $i$, and $\Delta t$ the sampling interval.

\begin{figure}[H]
\centering
\includegraphics[width=0.9\linewidth]{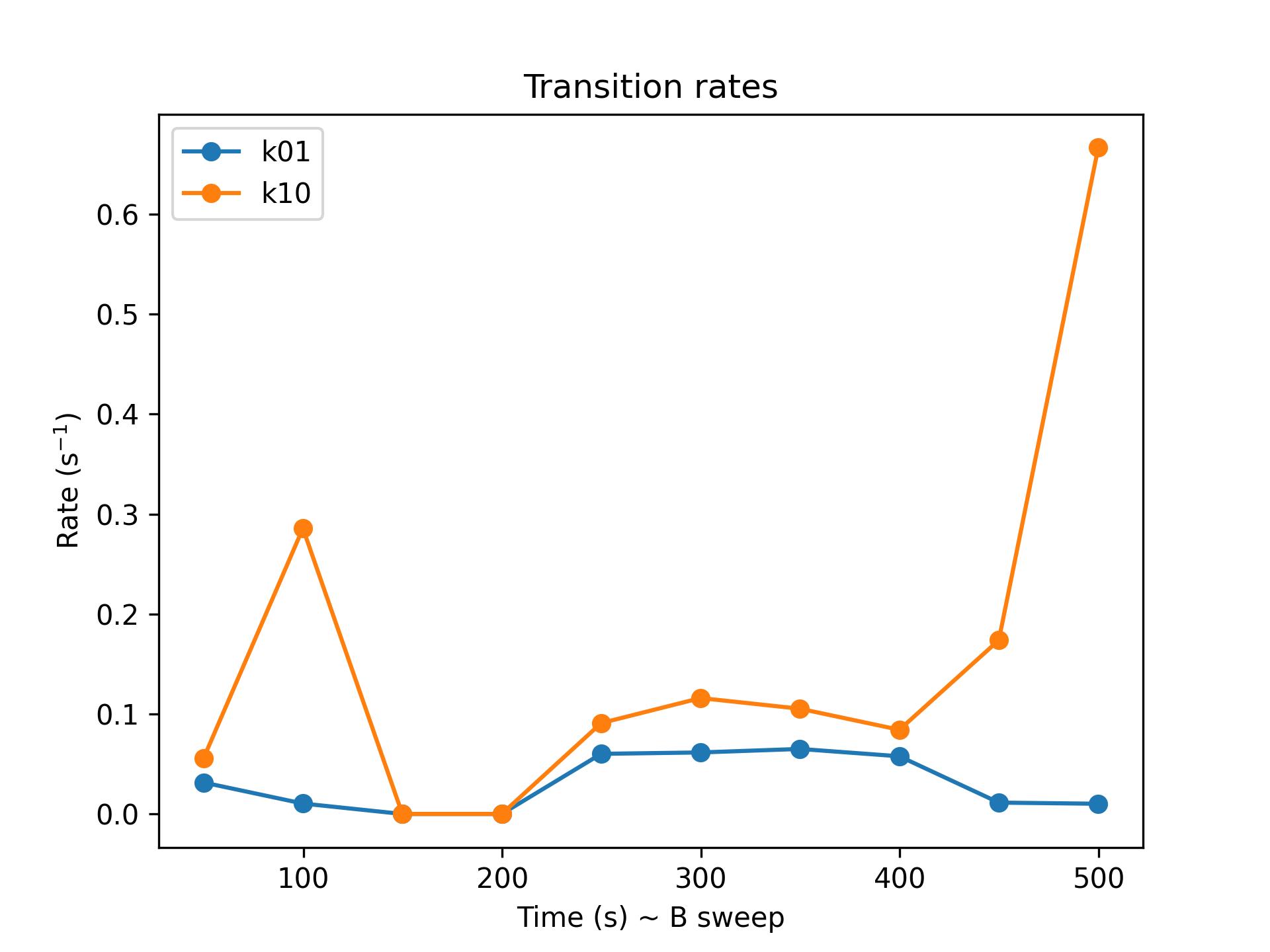}
\caption{Magnetic-field dependence of transition rates extracted from time-domain classification. The field primarily modulates the occupation probabilities of the transverse states rather than their voltage values.}
\label{fig:S5}
\end{figure}

The resulting rates are strongly asymmetric, with the high-voltage state exhibiting significantly longer dwell times. Typical lifetimes range from $\sim$20 s for the low state to $\sim$70 s for the high state. The corresponding field dependence of the extracted transition rates is shown in Fig.~\ref{fig:S5}.

\begin{figure}[H]
\centering
\includegraphics[width=0.9\linewidth]{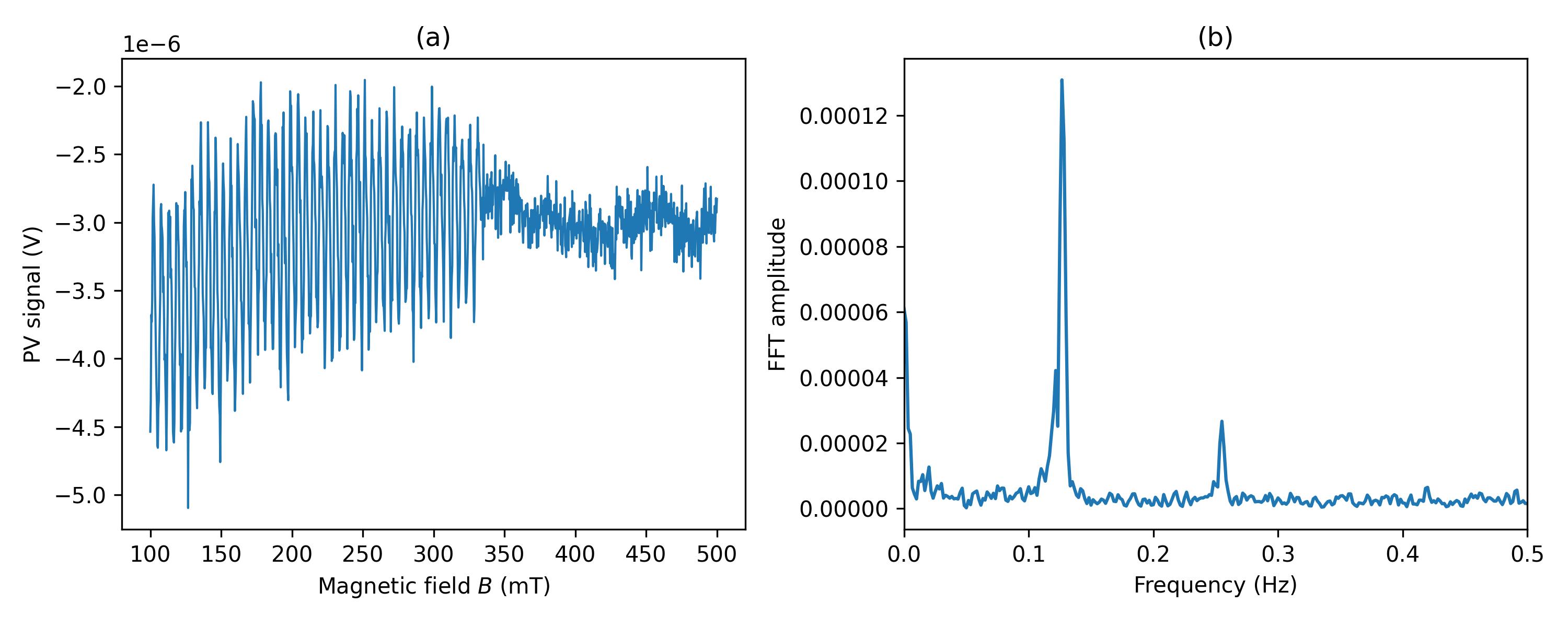}
\caption{(a) Photovoltaic (PV) signal measured simultaneously with $V_{xy}$ during a linear magnetic-field sweep at $T=19.9^\circ$C. 
(b) FFT amplitude spectrum of the PV signal, revealing a dominant narrow-band component near $f\approx0.13$~Hz. The absence of corresponding narrow-band structure in $V_{xy}$ indicates distinct dynamical responses of the PV and transverse channels.}
\label{fig:S6}
\end{figure}

Figure~\ref{fig:S6} provides an independent view of the photovoltaic response. The PV channel exhibits a coherent oscillatory mode with a well-defined characteristic frequency, in contrast to the broadband, switching-dominated dynamics observed in the transverse voltage.

\subsection{Field-resolved voltage states and photovoltaic response}
\noindent\textbf{Experimental conditions:} Hydrated genomic DNA (500~ng~$\mu$L$^{-1}$, 5~$\mu$L), 0.7$\times$0.7~cm$^{2}$ substrate, $T=19.4^{\circ}$C, RH $\sim24\%$, dark, continuous linear magnetic-field sweep.

\begin{figure}[H]
\centering
\includegraphics[width=0.9\linewidth]{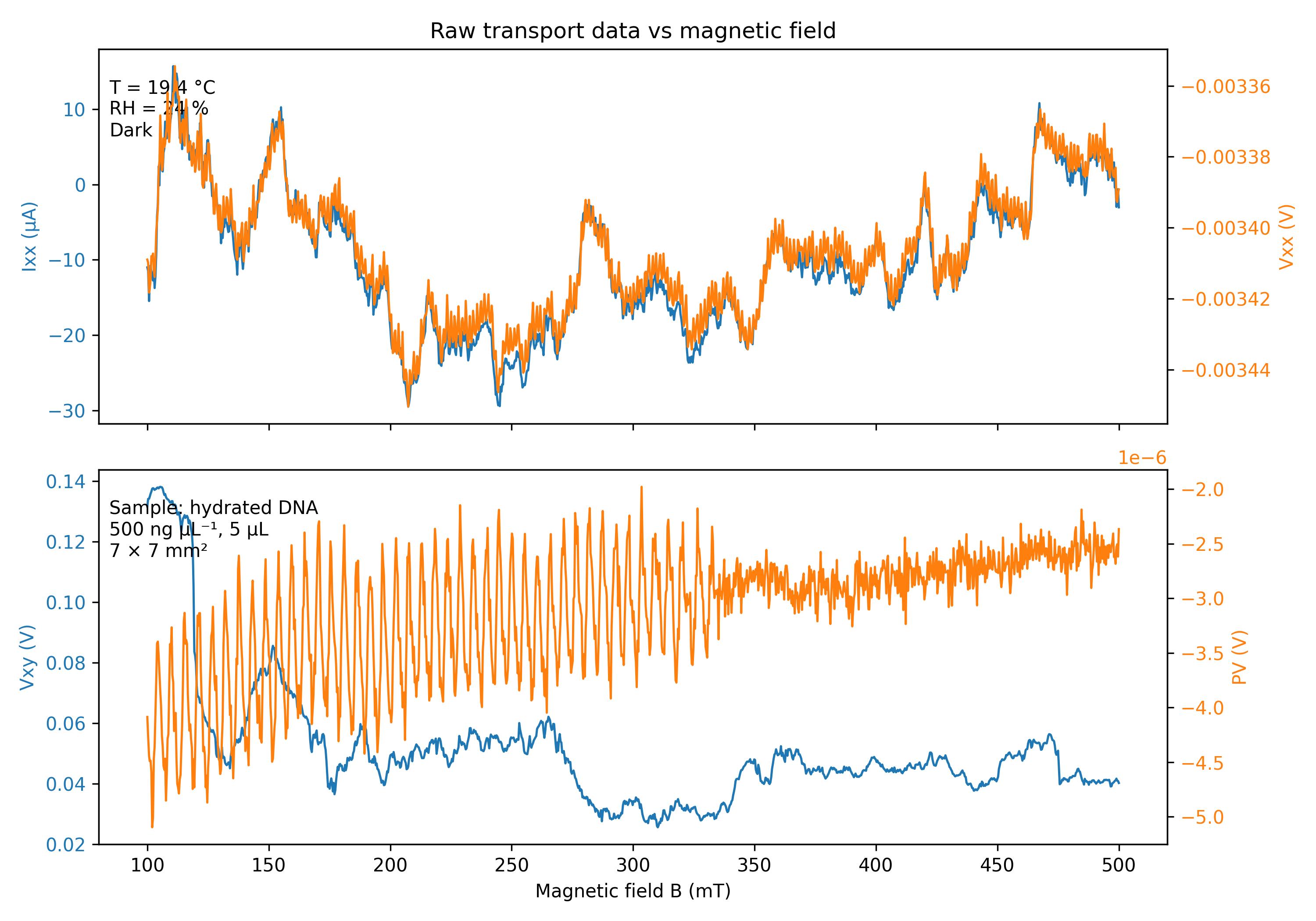}
\caption{Raw transport signals as a function of magnetic field at $T=19.4^\circ$C.
Upper panel: longitudinal current $I_{xx}$ (via shunt) and longitudinal sample voltage $V_{xx}$.
Lower panel: transverse voltage $V_{xy}$ and simultaneously recorded photovoltaic (PV) signal.
A sharp collapse of $V_{xy}$ near $B\approx120$~mT is followed by extended voltage plateaus, while $I_{xx}$ and $V_{xx}$ remain smooth and ohmic.}
\label{fig:SII_raw}
\end{figure}

Figure~\ref{fig:SII_raw} shows the raw transport signals
recorded during a magnetic-field sweep from approximately
100 to 500~mT.

Several robust experimental features are immediately apparent:

\begin{enumerate}
\item The transverse voltage exhibits an abrupt, large-amplitude transition
($\Delta V_{xy}\sim 80$--90~mV) at a well-defined magnetic field.
\item Beyond this threshold, $V_{xy}$ does not vary continuously with field but instead
forms extended plateaus with weak magnetic-field dependence.
\item The longitudinal transport channel remains smooth
and ohmic throughout the sweep, arguing against an abrupt
longitudinal-current instability as the origin of the
transverse-voltage steps.
\end{enumerate}

\begin{figure}[H]
\centering
\includegraphics[width=0.9\linewidth]{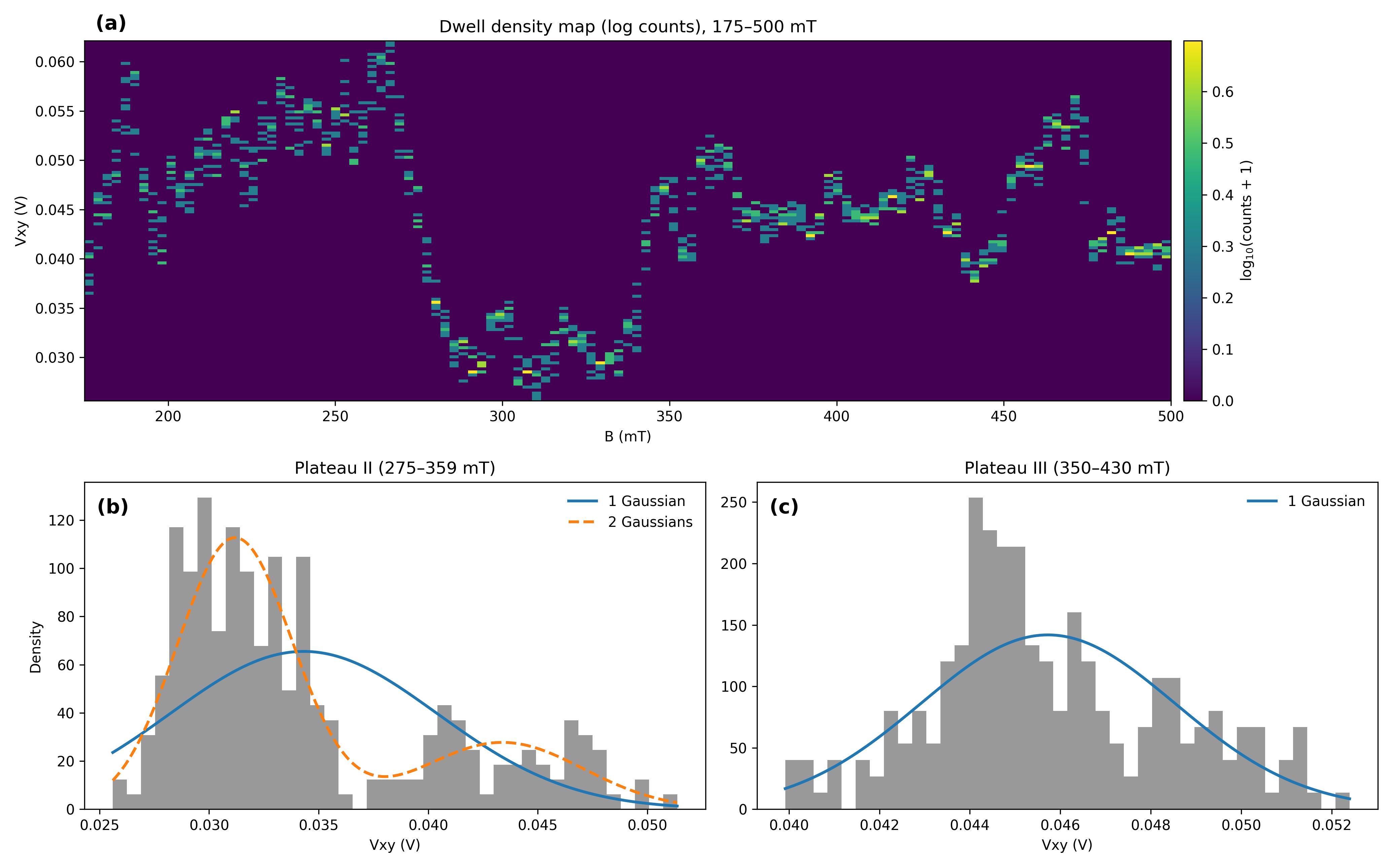}
\caption{Statistical evidence for metastable transverse attractors.
(a) Dwell-density map of $V_{xy}$ versus magnetic field (logarithmic counts), revealing discrete residence bands.
(b) Voltage histogram within Plateau~II (275--359~mT), showing a bimodal distribution described by a two-Gaussian mixture, indicating attractor coexistence.
(c) Voltage histogram within Plateau~III (350--430~mT), exhibiting a unimodal Gaussian distribution corresponding to a re-stabilized single attractor.}
\label{fig:SII_attractors}
\end{figure}

To characterize the nature of the plateau states, the time series was analyzed using
a dwell-density representation and voltage histogram statistics, as summarized in
Fig.~\ref{fig:SII_attractors}.

The bimodal voltage distributions identify two preferred levels;
switching between them is examined in the time-domain analysis above.
 
The dwell-density map demonstrates that the transverse voltage accumulates at discrete
values over extended magnetic-field intervals. These residence bands correspond to
long-lived metastable states rather than transient fluctuations.

Quantitative statistical analysis yields the following sequence of regimes:

\begin{itemize}
\item Single-attractor regime (Plateau~I):
The voltage distribution is unimodal and well described by a single Gaussian,
indicating stable phase locking within one attractor.

\item Attractor-competition regime (Plateau~II):
The distribution becomes bimodal, requiring a two-Gaussian mixture model.
This behavior reflects stochastic switching between two coexisting metastable states,
manifesting experimentally as telegraph-like voltage fluctuations.

\item Recondensed regime (Plateau~III):
At a higher magnetic field, the distribution returns to a unimodal form, signaling
condensation into a new stable attractor.
\end{itemize}

The observed sequence
\[
\text{single attractor}
\;\rightarrow\;
\text{attractor coexistence}
\;\rightarrow\;
\text{single attractor}
\]
is the characteristic fingerprint of a driven, dissipative phase system possessing a
compact order-parameter manifold.

\subsection{Simultaneous transverse-voltage and photovoltaic response}

\noindent\textbf{Experimental conditions:} Hydrated genomic DNA (500~ng~$\mu$L$^{-1}$, 5~$\mu$L), 0.7$\times$0.7~cm$^{2}$ substrate, $T\approx19.8$--19.6$^{\circ}$C, RH $\sim23\%$, dark, simultaneous transverse-voltage and photovoltaic acquisition during linear magnetic-field sweep.

\begin{figure}[H]
  \centering
  \includegraphics[width=\linewidth]{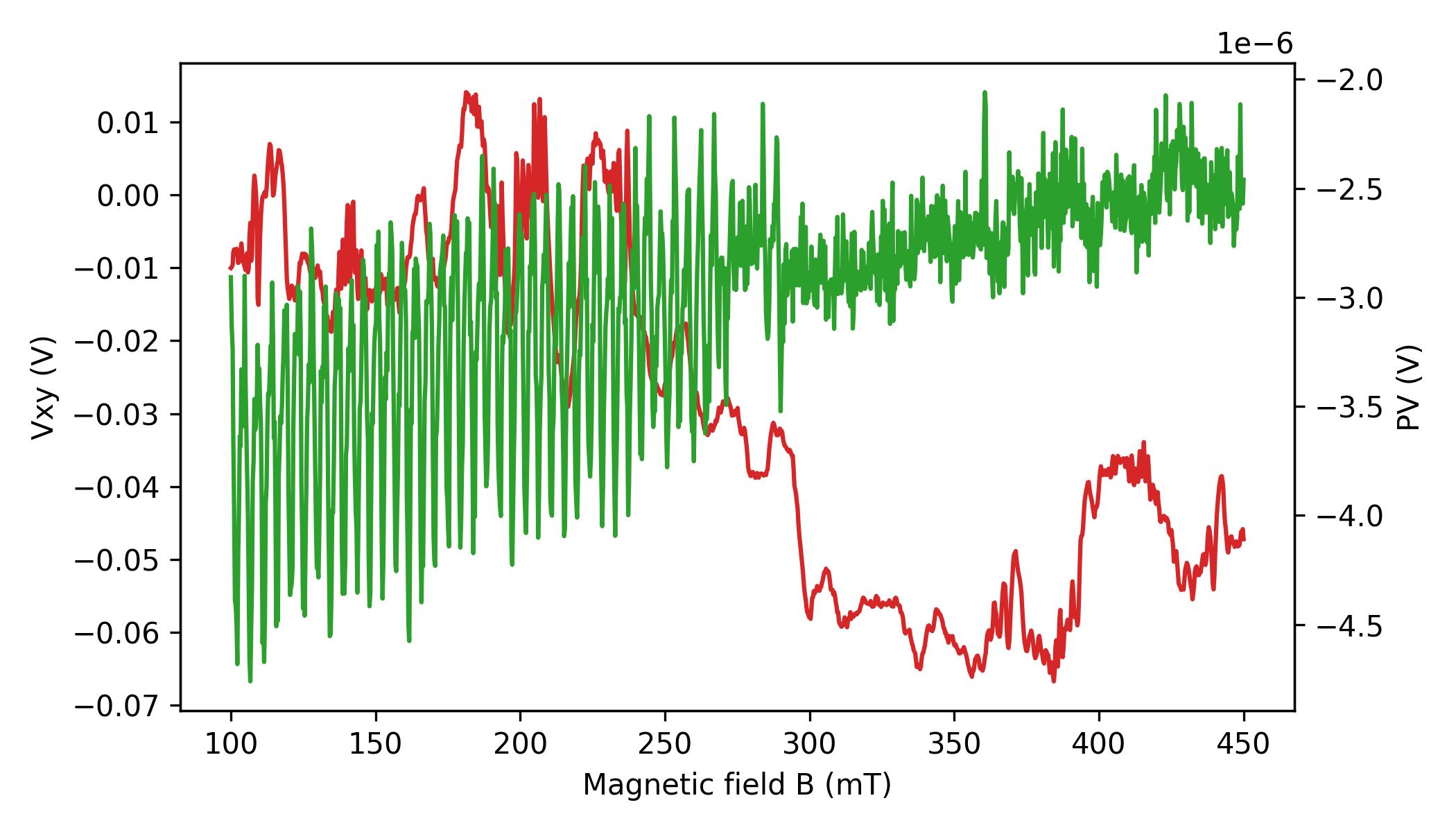}
  \caption{Simultaneous transverse voltage $V_{xy}$ (red, left axis) and photovoltaic signal PV (green, right axis) measured during a linear magnetic-field sweep at $T\approx19.8$--19.6$^\circ$C. Both channels exhibit a pronounced reorganization in the range $B\approx250$--300~mT, where $V_{xy}$ collapses into a deep negative basin while the PV oscillatory regime reorganizes. Sliding-window analysis (not shown) indicates anticorrelation between PV and $V_{xy}$ across the transition, consistent with a coupled collective response.}
  \label{fig:vxy_pv_threshold}
\end{figure}

The magnetic-field sweep covered approximately 100--450~mT,
with acquisition time mapped linearly to magnetic field. The PV signal was acquired on a physically separate detection channel, without electrical connection to the transport circuitry, providing an independent probe of the sample response. Figure~\ref{fig:vxy_pv_threshold} shows the resulting $V_{xy}$ (red) and PV (green) signals as functions of magnetic field. Both observables exhibit a pronounced change in behavior within the same magnetic-field interval ($B \approx 250$--300~mT): $V_{xy}$ undergoes a sharp transition into a negative basin (minimum $\sim -60$~mV), while the PV channel simultaneously reorganizes from a strongly oscillatory regime into a quieter biased state. A sliding-window cross-correlation analysis indicates that the PV signal becomes anticorrelated with $V_{xy}$ across this transition. The longitudinal voltage $V_{xx}$ and the corresponding
current $I_{xx}$ remain smooth and approximately ohmic
throughout the sweep (not shown), arguing against an abrupt
longitudinal-current instability as the source of the transition.
The coincident reorganization in two electrically separate
detection channels supports a common response associated
with the hydrated DNA system, while possible influences
from their shared experimental environment remain to be assessed.

\subsection{Robustness and reproducibility}

Magnetic-field thresholds followed by stepwise transverse responses have been observed reproducibly across multiple measurements performed under comparable ambient conditions.
While the absolute values and spacings of individual plateaus vary moderately between runs (0.8--1.0~mV), the qualitative organization into discrete metastable states separated by fluctuation-dominated transition regions is consistently preserved.

Related signatures of discrete and oscillatory behavior, including field-dependent oscillations and dense low-temperature switching patterns, have been reported in earlier measurements on similar hydrated DNA systems and are documented in the arXiv version of that work.
Across these datasets, the integer ordering of states and the occurrence of telegraph-like switching near transition boundaries remain robust, even as noise levels and microscopic conditions change.

Separate control measurements provide additional context
for the sample dependence of the response
\cite{Pietruszka2026ArXiv}. The distilled-water control
lacked the pronounced voltage structures observed in
hydrated DNA, while a Cu-coated substrate produced
a transverse response of opposite sign. These comparisons
support an association of the structured response with
the DNA-containing sample and complement the simultaneous
transverse-voltage and photovoltaic measurements
presented here.

Forward and reverse magnetic-field sweeps were compared
in a separate control study, where the principal voltage
features occurred at approximately similar field positions
\cite{Pietruszka2026ArXiv}. The measurements presented here
did not include a field-cycling protocol, so hysteresis
of the present staircase response was not assessed.

\section{Phenomenological model}

The voltage plateaus and telegraph-like switching motivate
a phenomenological description in terms of metastable states
and transitions between them. Related measurements in hydrated
DNA also exhibit sustained Hilbert-phase winding, with successive
circuits of the complex analytic trajectory around the origin
\cite{Pietruszka2026ArXiv}. These observations motivate consideration
of both switching and running-phase dynamics within a broader
collective-polarization framework. Here, the model focuses on
the metastable response documented in the present measurements.
The Hilbert phase reconstructed from the voltage is a
signal-derived observable; its relation to the phase of a
collective polarization field remains a physical hypothesis.

\begin{figure}[H]
    \centering
    \includegraphics[width=0.80\linewidth]{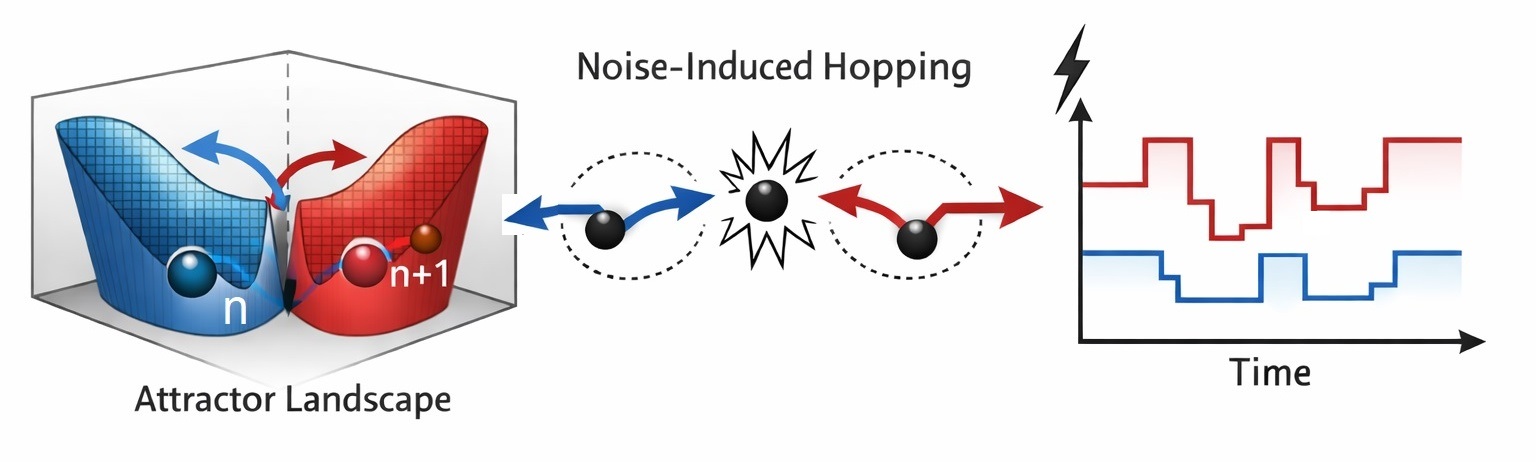}
\caption{
Phenomenological schematic of switching between two metastable
states in an effective phase potential $U(\theta;B)$.
Local minima represent preferred states separated by a barrier.
Stochastic transitions between the minima provide a possible
description of telegraph-like voltage switching and bimodal
voltage distributions. The corresponding overdamped Langevin
and Fokker--Planck equations are given in
Eqs.~\ref{eq:phase} and \ref{eq:phase_FP}. Schematic illustration generated using generative AI under the author's direction.
}
    \label{fig:S1}
\end{figure}

\begin{figure}[H]
    \centering
    \includegraphics[width=0.95\linewidth]{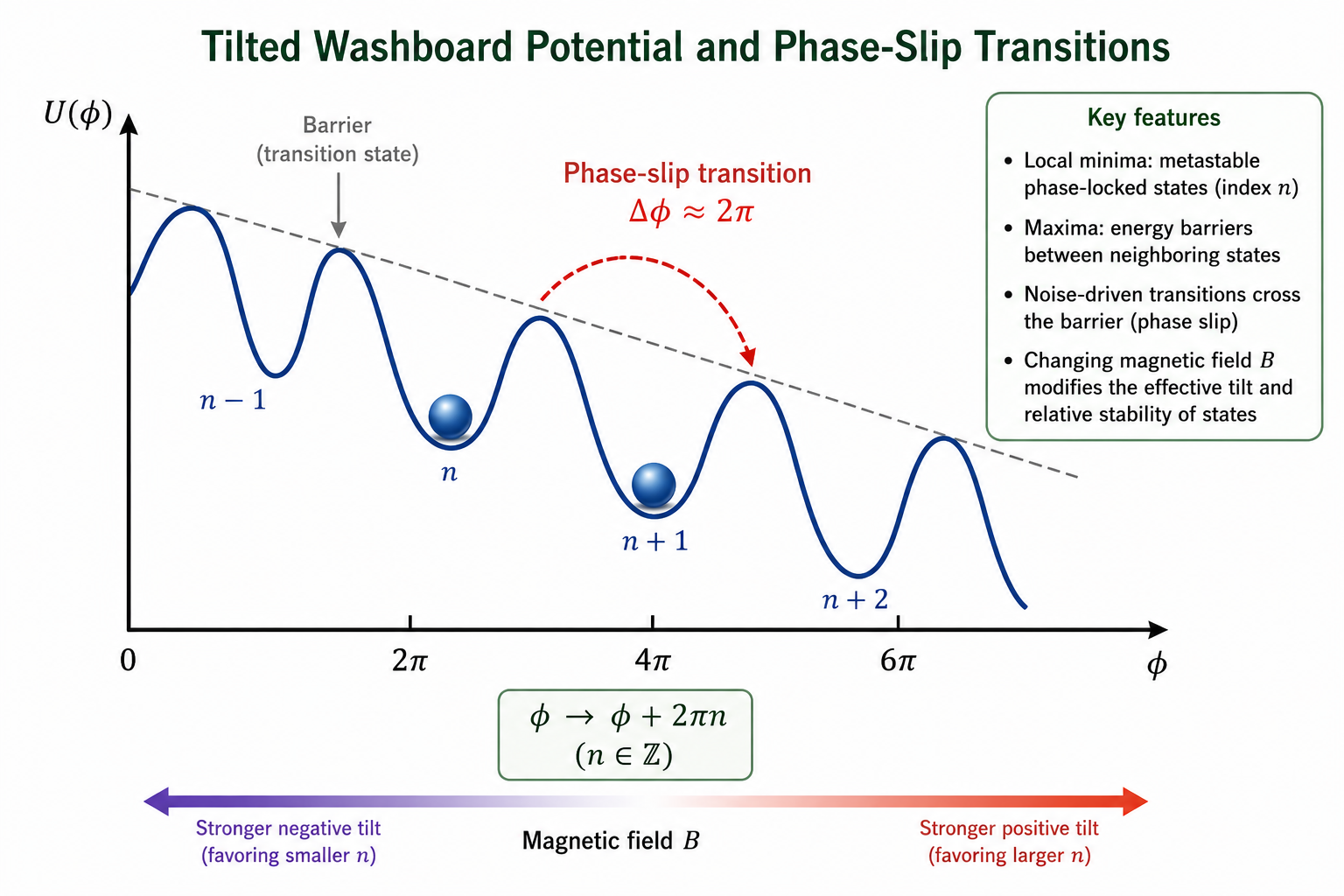}
    \caption{
Tilted washboard potential illustrating metastable phase attractors and phase-slip transitions in the collective polarization field. Local minima correspond to phase-locked states labeled by an integer index $n$, while maxima represent barriers separating neighboring states. Thermal and environmental fluctuations can induce noise-assisted phase-slip transitions of order $\Delta\phi \approx \pm 2\pi$ between adjacent attractors. Variation of the magnetic field $B$ modifies the effective tilt of the potential and thereby controls the relative stability of the metastable phase sectors.
Schematic illustration generated using generative AI
under the author's direction.}
    \label{fig:washboard}
\end{figure}

The phenomenological model is illustrated in
Fig.~\ref{fig:S1}, showing switching between neighboring
metastable states, and Fig.~\ref{fig:washboard}, showing
the tilted periodic potential.

We introduce a complex collective field
\begin{equation}
\psi(t)=|\psi(t)|e^{i\theta(t)}.
\end{equation}
Here, $|\psi|$ represents the amplitude of a proposed collective
polarization mode, and $\theta$ is its phase, defined modulo
$2\pi$. The measured voltage $V_{xy}$ provides an electrical
readout whose relation to $\psi$ depends on the sample,
electrodes, and measurement circuit. A quantitative mapping
between this field and the voltage-derived Hilbert phase
has not been established.

Under ambient conditions, the dynamics are strongly dissipative, and an overdamped Langevin equation may describe the slow evolution of the collective phase,
\begin{equation}
\gamma \dot{\theta}(t)
=
- \frac{\partial U(\theta;B)}{\partial \theta}
+ \xi(t),
\label{eq:phase}
\end{equation}
where $\gamma$ is an effective damping coefficient, $U(\theta;B)$ is an effective field-dependent phase potential, and $\xi(t)$ represents thermal and environmental noise. 

For the stochastic model, we assume Gaussian white noise,
\begin{equation}
\langle\xi(t)\rangle=0,
\qquad
\langle\xi(t)\xi(t')\rangle
=2\gamma D\,\delta(t-t'),
\end{equation}
Here, $D$ parameterizes the effective stochastic forcing
of the proposed collective phase.
The measured voltage fluctuations may contain contributions
from sample dynamics, environmental disturbances, and
instrumental noise; their variance is not a direct measurement
of $D$ without a calibrated phase-to-voltage relation.
The white-noise assumption is phenomenological, and $D$
was not independently controlled or calibrated in these experiments.

Under this assumption, Eq.~\ref{eq:phase} corresponds to the
Fokker--Planck equation for the phase probability density
$P(\theta,t)$,
\begin{equation}
\partial_t P(\theta,t)
=
\frac{1}{\gamma}\partial_\theta
\left[
\left(\partial_\theta U\right)P(\theta,t)
\right]
+
\frac{D}{\gamma}\partial_\theta^2 P(\theta,t).
\label{eq:phase_FP}
\end{equation}
The drift term describes evolution in the effective potential,
while the diffusion term represents stochastic fluctuations.
Together, these equations provide a phenomenological description
of residence near preferred states and transitions between them.

A periodic phase potential provides a phenomenological picture
of trapping and noise-assisted transitions. With an effective
drive, a tilted potential can also support running-phase
dynamics, in which the unwrapped phase advances through
successive periods. These regimes offer a common framework
for discussing switching and sustained winding.
However, a $2\pi$ advance returns the compact phase to the
same angular value; distinct voltage plateaus therefore require
a physical distinction between collective configurations
beyond this phase periodicity alone.

The model offers a qualitative description of metastable
residence and switching, rather than a quantitative derivation
of the voltage staircase. Its stochastic formulation is given
in Eqs.~\ref{eq:phase} and \ref{eq:phase_FP}; the experimental
voltage distributions and Bayesian model comparison are
presented in Figs.~\ref{fig:attractors} and \ref{fig:bic_overlay}.

The effective phase dynamics can be visualized as motion in a tilted washboard potential (Fig.~\ref{fig:washboard}). In this picture, local minima represent metastable phase-locked attractors labeled by an integer index $n$, while barriers between neighboring minima correspond to transition states. Noise-assisted barrier crossing produces phase-slip events of order $\Delta\phi \approx \pm 2\pi$, in the unwrapped phase representation.

\section{Topological Phase--Field Mechanics at Room Temperature}

The phase-field interpretation provides a possible connection
between metastable switching and continuous phase evolution.
Related voltage measurements motivate a collective-polarization
description \cite{Pietruszka2026BioSystems,Pietruszka2026ArXiv}.
However, establishing topologically distinct collective states
requires a physical definition of the field, its spatial
configuration, and the relevant boundary conditions.
The voltage distributions and reconstructed Hilbert phase
alone do not establish this spatial topology.

\subsection{Scope of the phase-field interpretation}

The proposed collective field is a phenomenological variable.
Its microscopic origin, magnetic coupling, and quantitative
relation to the measured voltage remain to be established.
Temporal winding of the reconstructed Hilbert phase characterizes
the voltage dynamics, whereas spatial winding of a physical
polarization field would require additional evidence.
Accordingly, the present model does not derive the plateau
spacing or establish topological protection.

Phase-based models of collective dynamics have established
precedents in the XY model, vortex physics, Josephson-junction
arrays, and driven dissipative systems
\cite{Kosterlitz1973,Nelson1977,Fazio2001,Sieberer2016}.
These examples provide conceptual context for the proposed
description; they do not establish the same microscopic
mechanism or topological structure in hydrated DNA.

\subsection{Topological--Wave Dualism of a Collective Polarization Field}

Discrete voltage plateaus and oscillatory responses reveal
two complementary aspects of the measured dynamics:
residence near preferred macroscopic levels and continuous
temporal evolution. Sustained Hilbert-phase winding in related
measurements further illustrates the continuous aspect
\cite{Pietruszka2026ArXiv}. We use the term
\emph{topological--wave dualism} as a proposed interpretation
linking wave-like collective dynamics with discrete metastable
configurations. Here, ``dualism'' refers to these complementary
dynamical features, while their common physical origin and
topological organization remain to be established.

Within this interpretation, a smooth collective phase field
$\phi(\mathbf r,t)$ could support oscillations and phase
correlations. If the field remains well defined along a closed
spatial contour, its phase can accumulate an integer winding,
consistent with $\pi_1(U(1))=\mathbb Z$.
This provides a possible connection between continuous local
dynamics and discrete global configurations. The temporal
Hilbert-phase winding reconstructed from voltage traces
characterizes the signal evolution; it does not by itself
identify such spatial winding sectors.

\subsubsection{Wave-like phase dynamics and temporal interference}

We assume that the system supports a coherent collective polarization mode
described by a complex order parameter
\begin{equation}
\psi(\mathbf r,t)
=
|\psi(\mathbf r,t)|\,e^{i\phi(\mathbf r,t)},
\end{equation}
where $\phi(\mathbf r,t)$ denotes the phase of the proposed
collective mode. The longitudinal and transverse voltages
may contain contributions from its dynamics, with different
couplings to the electrodes and measurement circuit.
Their quantitative dependence on the collective phase
has not been established.

The temporal interference map,
\begin{equation}
S(t,\Delta t)
=
V_{xx}(t)
+
V_{xy}(t+\Delta t),
\end{equation}
exhibits alternating regions of signal reinforcement and
cancellation, with an estimated phase offset of approximately
$200^\circ$. This pattern characterizes the relative timing
of the two voltage traces and is consistent with correlated
oscillatory components. The numerical superposition provides
a useful visualization of their phase relationship; establishing
physical interference of a common collective mode requires
additional evidence.

A possible phenomenological parameterization of the relative
phase is
\begin{equation}
\Delta\phi(\omega)
\simeq
\phi_0
+
\arctan(\omega\tau)
+
\omega\Delta t,
\end{equation}
where $\phi_0$ is a baseline phase offset, $\tau$ an effective
relaxation time, and $\Delta t$ the sequential sampling delay.
This expression illustrates possible contributions to the
measured phase relationship; these contributions have not
been separately determined in the present analysis.

\subsubsection{Topological sectors and particle-like response}

For a spatial phase field that is continuous and well defined
along a closed contour, the accumulated phase satisfies
\begin{equation}
\oint
\nabla \phi \cdot d\boldsymbol{\ell}
=
2\pi n,
\qquad
n \in \mathbb Z,
\end{equation}
where $n$ is the spatial winding number. Such winding could
distinguish collective configurations, provided the physical
field and the relevant contours are identified. The present
voltage measurements do not directly determine this spatial index.
The measured voltage plateaus identify preferred macroscopic
response levels. Their distributions are approximately unimodal
within selected plateau intervals and become bimodal near
some transition boundaries. These observations support a
description in terms of metastable states, but do not determine
a spatial winding number or establish a voltage--winding
response law.

\subsection{Analogy with coupled phase arrays}

Josephson-junction arrays provide a useful reference for
coupled phase dynamics, frustration, metastability, and
phase-slip transitions \cite{Fazio2001}.
An analogous description of hydrated DNA could involve
coupled mesoscopic polarization domains. Applying such a
model quantitatively would require identifying these domains,
their coupling, and the mechanism by which the magnetic
field influences their phases. The analogy therefore motivates
possible dynamical models rather than establishing a direct
mapping or a voltage--winding relation.

\subsection{Plateau spacing from a phase--voltage conjugacy}

To connect the integer winding number to the observed voltage staircase, we
introduce a minimal phenomenological conjugacy between the transverse polarization voltage and the collective phase dynamics. At the level of coarse-grained response, a
voltage-like observable may be taken to be proportional to the time derivative
of an effective phase variable,
\begin{equation}
V_{xy}(t)
=
\kappa\,\partial_t \varphi(t),
\label{eq:pol_josephson}
\end{equation}
where $\varphi(t)$ denotes the phase drop conjugate to the measured transverse
polarization response and $\kappa$ is a system-dependent conversion constant.
Equation~(\ref{eq:pol_josephson}) is formally analogous to the Josephson
phase--voltage relation, but here applies to a collective polarization mode in
a driven, dissipative medium rather than to an electronic superconducting
condensate.

In the steady states relevant to the plateau regime, the system relaxes into
phase-locked attractors characterized by an integer winding number
\begin{equation}
n
=
\frac{1}{2\pi}
\oint_{\mathcal C}
\nabla\phi \cdot d\boldsymbol{\ell}
\in \mathbb Z ,
\end{equation}
such that the gauge-invariant phase accumulation across the transverse
measurement contour is quantized,
\begin{equation}
\Delta\varphi_n = 2\pi n .
\label{eq:phase_quant}
\end{equation}
A phase-slip event corresponds to a vortex crossing the contour and changes
the winding by $\Delta n=\pm1$, implying a quantized change of the phase drop
\begin{equation}
\Delta(\Delta\varphi)
=
\Delta\varphi_{n+1}-\Delta\varphi_n
=
2\pi .
\end{equation}

Within a given topological sector, phase locking implies a fixed mean
phase-advance rate $\langle\partial_t\varphi\rangle_n$. Assuming that the drive selects mean phase-advance rates
$\langle\partial_t\varphi\rangle_n=n\Omega$,
the phase--voltage relation
in Eq.~(\ref{eq:pol_josephson}) yields
\begin{equation}
\langle V_{xy}\rangle_n
=
\kappa\,\langle\partial_t\varphi\rangle_n
\equiv
\alpha\, n ,
\end{equation}
where we define $\alpha \equiv \kappa\,\Omega$ and
\begin{equation}
\Omega
\equiv
\langle\partial_t\varphi\rangle_{n+1}
-
\langle\partial_t\varphi\rangle_n
\end{equation}
is the approximately constant phase-advance increment selected by the external
drive and dissipation.

Consequently, the plateau spacing is
\begin{equation}
\Delta V
=
\langle V_{xy}\rangle_{n+1}
-
\langle V_{xy}\rangle_n
=
\alpha
\end{equation}
Within these assumptions, a regularly spaced voltage staircase
can be represented through (i) the existence
of a phase--voltage conjugacy for the collective polarization field and (ii)
topological quantization of the phase winding, without invoking spectral
quantization of microscopic energy levels.

Equation~(\ref{eq:pol_josephson}) should therefore be interpreted as a generic
response relation for a driven collective mode. The conversion constant
$\kappa$ is system-specific and is not expected to coincide with the
superconducting value $\hbar/2e$.

\subsection{From Fr\"ohlich Coherence to Topological Attractors}

Following the proposal of Fr\"ohlich~\cite{Frohlich1968,Frohlich1970}, driven and dissipative hydrated matter can undergo nonlinear mode selection, whereby energy injected into many coupled polar degrees of freedom concentrates into a small number of long-wavelength collective modes once a threshold is exceeded. This Fr\"ohlich-type mechanism provides a general route to macroscopic coherence in warm, lossy biological media mediated by water dipoles and hydrogen-bond networks. In the proposed interpretation, magnetic-field excitation and electrical bias promote collective polarization dynamics
and the selection of ordered macroscopic states, consistent with recent experimental observations in hydrated DNA~\cite{Pietruszka2025Frohlich}. Near the boundaries between such states, enhanced fluctuations give rise to telegraph switching, revealing real-time competition between neighboring metastable configurations. Subsequent relaxation selects a single attractor, erasing microscopic pathway information and leaving only an integer-valued macroscopic order parameter. This behavior is naturally described as a dissipative projection of continuous polarization dynamics onto discrete topological sectors of an emergent $U(1)$ phase field, with phase-slip events mediating transitions between attractors. In this sense, the observed plateaus and telegraph switching can be interpreted as signatures of Fr\"ohlich-type collective organization followed by stabilization of metastable configurations, providing a minimal physical realization of the long-range ordering mechanisms discussed in dissipative many-body approaches to biological matter~\cite{Vitiello2004}.

In dissipative many-body field theory, the doubling of degrees of freedom introduced by Vitiello~\cite{Vitiello2001,Vitiello2004}, following the thermo-field dynamics framework of Umezawa~\cite{Umezawa1993}, provides a natural description of irreversible dynamics, with distinct functional states corresponding to inequivalent vacua stabilized by environmental coupling. A complementary geometric interpretation of the present system
can be formulated within the proposed phase-field picture: a collective $U(1)$ polarization phase field decomposes into integer-labeled topological sectors, each representing a metastable attractor. Telegraph switching reflects transient competition between neighboring sectors, during which phase-slip events temporarily reconnect otherwise disconnected configurations. Relaxation then stabilizes a single winding number, erasing microscopic pathway information.
Thus, Vitiello's doubled Hilbert-space formulation and the
proposed $U(1)$ phase-field picture offer complementary
perspectives on dissipative state selection: the former
emphasizes system--environment coupling, while the latter
explores the geometric organization of collective configurations.

The observation of well-separated transverse voltage plateaus and telegraph switching with characteristic nearest-neighbor spacing provides direct evidence for discrete collective states in a driven hydrated DNA system at ambient conditions (see Fig.~\ref{fig:SII_attractors}). The bimodal voltage distributions, supported by representative
model comparisons with $\Delta\mathrm{BIC}\approx100$, identify
two preferred response levels. Together with the time-domain
switching, they support a description in terms of metastable
attractors and fluctuation-assisted transitions. Within the proposed phase-field interpretation, these states
are modeled as integer-labeled polarization configurations arising from topologically distinct winding sectors of a collective $U(1)$ order parameter, with transitions mediated by phase-slip events. The large separation between attractors relative to intra-plateau noise indicates deep collective phase basins, thereby enabling robust room-temperature readout. More broadly, these results establish hydrated DNA as a soft-matter platform exhibiting emergent macroscopic quantization through collective phase organization, suggesting a general mechanism for topologically structured dynamics in driven biological media. Taken together, these results support topological phase-field mechanics as a minimal, experimentally grounded framework for discrete macroscopic response in warm, driven, and dissipative matter, suggesting that robust ambient-condition quantization can emerge from collective phase topology rather than microscopic energy spectra.

\section{Discussion}

The measurements reveal discrete transverse-voltage levels
and telegraph-like switching in hydrated DNA under magnetic-field
excitation at ambient temperature. Voltage distributions and
Bayesian model comparisons support the identification of
preferred response levels, while the time traces resolve
transitions between them. These findings extend the earlier
characterization of the ohmic--polarization crossover
\cite{Pietruszka2026BioSystems} by resolving the metastable
structure of the post-threshold response. Related bistability
during cooling at fixed magnetic field provides a complementary
experimental context \cite{Pietruszka2026BistableDNA}.

The phase-field picture connects these observations with
a broader range of collective dynamics. Metastable residence
and fluctuation-assisted switching can be represented by
motion near minima of an effective potential, whereas sustained
phase evolution corresponds to a running regime. Related
measurements show successive circuits of the complex analytic
trajectory around the origin and sustained unwrapped
Hilbert-phase winding \cite{Pietruszka2026ArXiv}.
Together, these observations motivate a framework encompassing
both discrete response configurations and continuous phase
evolution, the complementary features described here as
topological--wave dualism.

Fluctuations may play an organizing role by enabling exploration
of competing metastable configurations and transitions between
them. This perspective connects the proposed switching picture
with the broader statistical physics of complex systems and
multiple states discussed by Parisi \cite{Parisi2023}.
The connection is conceptual.

Noise can also enhance metastable stability: studies of
switching potentials and Josephson junctions show that
mean escape times can depend non-monotonically on noise
intensity, with maxima at finite noise levels
\cite{Spagnolo2004NES,Augello2009}.
A distinct constructive effect, stochastic resonance,
has also been observed in biological signal detection
\cite{Spezia2008}.
These findings motivate controlled studies of fluctuations
in hydrated DNA. Here, noise intensity was not varied
systematically, so a lifetime--noise relation and
noise-enhanced stability remain questions for future investigation.

A related statistical signature occurs in memristive systems,
where noise-induced transient bimodality accompanies resistive
switching \cite{Agudov2021}.
Here, bimodal voltage distributions together with time-resolved
switching support occupation of two preferred response levels.
The analogy concerns switching statistics and motivates
residence-time analysis; it does not imply a common microscopic
mechanism or simultaneous equilibrium coexistence.

Within this interpretation, collective organization and
dissipative state selection provide a possible physical link
between the hydrated molecular environment and the macroscopic
voltage response. Fr\"ohlich's proposals for driven polar modes
\cite{Frohlich1968,Frohlich1975} and subsequent work on
nonequilibrium condensation \cite{Wang2022PRB} provide conceptual
context. Vitiello's dissipative formulation further emphasizes
the role of system--environment coupling in state selection
\cite{Vitiello2001,Vitiello2004}. These perspectives motivate
the proposed polarization description, while topological
approaches to driven nonlinear systems offer a complementary
geometric viewpoint \cite{Villa2025SciAdv}.
A possible physical realization is a set of metastable
collective polarization configurations involving DNA
and its hydration environment. In this picture, switching
reflects reorganization of the collective response, rather
than identical conformational changes of every DNA molecule.
The voltage measurements alone do not distinguish clustered,
extended, or other molecular arrangements; simultaneous
structural measurements would help identify the configurations
associated with the observed levels.

The experimentally reconstructed Hilbert phase characterizes
the voltage signal; identifying it with a physical polarization
phase requires a model of the electrical readout. Likewise,
a spatial winding index would require specification of the
collective field and its boundary conditions. The phase--voltage
relation developed above therefore represents a model assumption
through which regular phase-advance rates can produce regularly
spaced voltage levels. Here, ``quantized'' denotes the observed
discrete macroscopic response levels, while their proposed
topological origin remains an interpretation to be tested.

The recurrence of thresholds, preferred voltage levels, and
switching across the presented measurements motivates further
study of hydrated DNA as a medium for collective nonlinear
dynamics. Measurements combining controlled field protocols,
longer stationary recordings, and independent structural or
polarization probes could connect the macroscopic response
more directly to its physical configurations.

Beyond its relevance to soft-matter and biological physics,
the persistence of discrete metastable response levels under
ambient conditions suggests potential utility in noise-tolerant,
dissipative information-processing schemes.
More generally, the present findings motivate exploration
of phase topology as a possible organizing principle for
collective polarization dynamics far from equilibrium,
providing a potential route to discrete, robust macroscopic
states without requiring low-temperature energy quantization.

\section{Materials and Methods}

Genomic DNA was isolated from leaf tissue of barley
(\textit{Hordeum vulgare}, cultivar Sebastian) grown under
controlled conditions. DNA extraction from dried plant
material was performed using a modified micro-CTAB protocol
by Katarzyna Konopka. DNA quantity and purity were verified
spectrophotometrically using a NanoDrop ND-1000
spectrophotometer (Thermo Scientific).
Following isolation, the DNA was dissolved in 100~$\mu$L
of TE buffer (10~mM Tris--HCl, 1~mM EDTA, pH~8.3).
Before the experiments, this stock solution was diluted
with deionized water to a final DNA concentration of
500~ng~$\mu$L$^{-1}$. The measurements used hydrated,
double-stranded barley genomic DNA in the resulting
aqueous solution (pH~$\approx8$), containing residual
TE buffer from the stock solution.

% Experiments were performed on hydrated double-stranded barley genomic DNA dissolved in deionized water (pH $\approx 8$) at a concentration of 500~ng~$\mu$L$^{-1}$.
A droplet of volume 5~$\mu$L was deposited onto the substrate
and covered with a cover glass. Depending on lateral spreading,
the covered sample area was approximately 50--100~mm$^2$,
giving an estimated mean thickness of approximately
50--100~$\mu$m. At the stated concentration, each 5~$\mu$L droplet contained
2.5~$\mu$g of genomic DNA. The measurements therefore probed
the electrical response of a macroscopic ensemble of DNA
molecules, rather than an individual strand.
Electrical contacts were arranged to measure the transverse voltage response $V_{xy}$ under an applied dc bias.

A dc power supply delivered a constant longitudinal voltage $V_{xx}$ of 0.1~V with a current limit of 0.01~A.
The magnetic field was applied perpendicular to the sample plane using neodymium permanent magnets and adjusted manually in discrete steps (typical step duration of order seconds).
The magnetic-field induction was monitored using a calibrated teslameter (Leybold Didactic GmbH). Electrical measurements were performed using a Keithley DMM6500 precision digital multimeter (Tektronix/Keithley).
The measurements were carried out under ambient laboratory conditions at a temperature $T = 21.9^{\circ}$C and relative humidity of approximately 25\%.

The transverse voltage $V_{xy}$ was recorded continuously as a function of time during the magnetic-field sweep.
Time was subsequently mapped linearly to the magnetic field using the independently measured field endpoints.
The linear time-to-field mapping approximates the overall
sweep and does not resolve local variations during manual
field adjustment. Residence durations measured from the
time traces therefore characterize the response under
the applied sweep protocol; they are not intrinsic
fixed-field lifetimes. Plateau widths in field units
cannot be converted into such lifetimes without accounting
for the local sweep rate.
During manual adjustment of the magnetic field, brief transient voltage spikes were occasionally observed; these short-lived events do not correspond to stable dwell states and do not contribute to the plateau structure or dwell-density ridges discussed in the Results.

The phenomenological Langevin and Fokker--Planck formulations
are given in Eqs.~\ref{eq:phase} and \ref{eq:phase_FP}.

Complementary measurements using continuous magnetic-field sweeps
were performed at nearby ambient temperatures. Their voltage
statistics, spectral characteristics, and simultaneous transverse
and photovoltaic responses are presented in the Results.

\begin{figure}[H]
  \centering
  \includegraphics[width=0.72\linewidth]{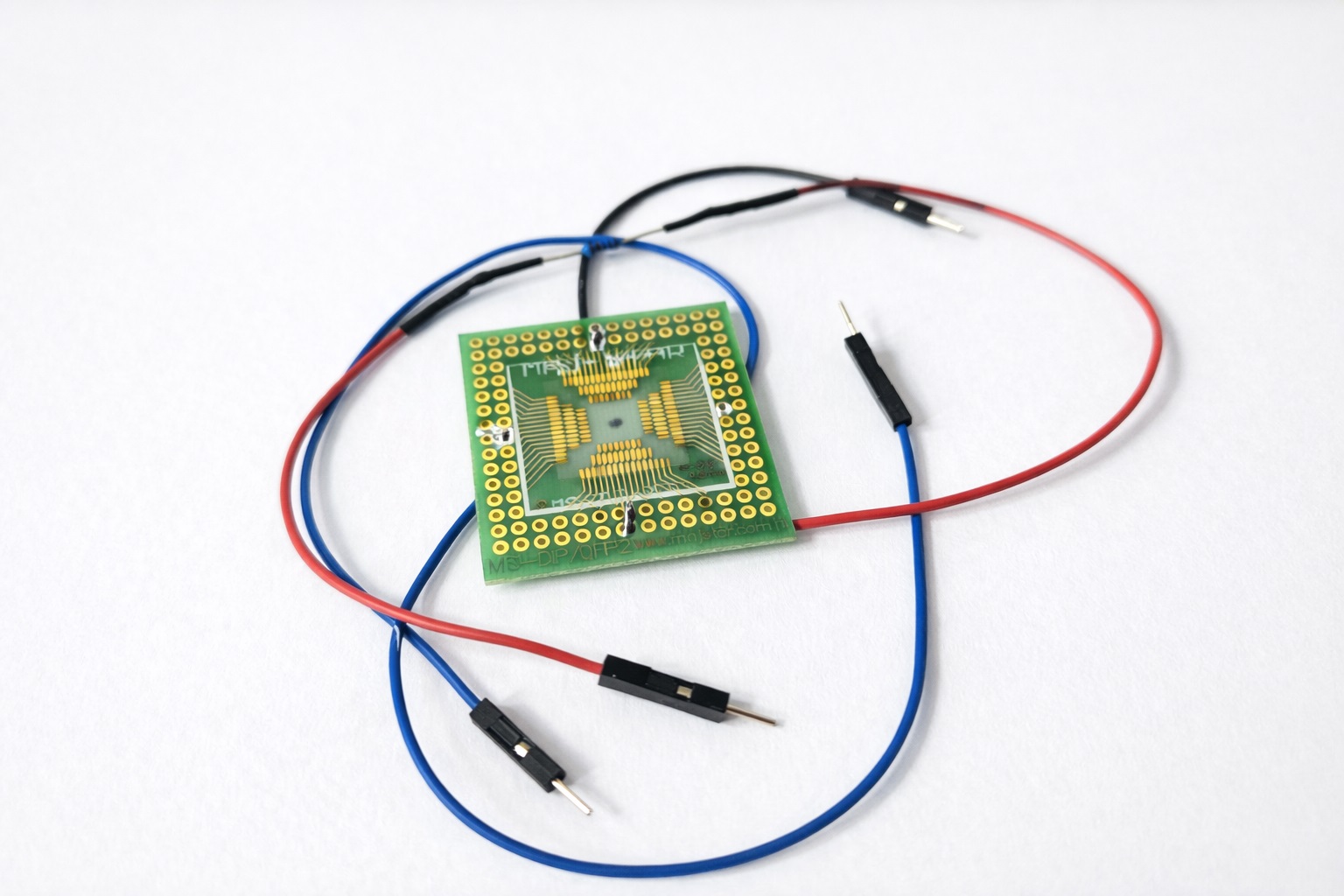}
  \caption{Photograph of the custom planar detector$_\text{MP}$ (here: in Greek-cross configuration) used for electrical measurements.
  The device consists of a Cu-coated planar substrate with a four-contact geometry mounted on an insulating support. A microscope glass coverslip placed over the active region mechanically flattens the liquid sample, enforcing a quasi-2D hydrated DNA layer.
  Electrical connections are made via shielded leads to minimize pickup and inter-channel cross-talk. A series shunt resistor visible in the supply line allows simultaneous determination of the longitudinal current $I_{xx}$ from the measured voltage drop across the resistor.}
  \label{fig:S10_detector}
\end{figure}

A photograph of the planar four-contact detector and coverslip configuration used to enforce a quasi-2D hydrated layer is provided in Fig.~\ref{fig:S10_detector}.

% Figure S10

\section*{Funding}
Not applicable.

\section*{Acknowledgements}

The author gratefully acknowledges Katarzyna Konopka for assistance with genomic DNA sample preparation used in this study.

\section*{Data availability}

The datasets generated and analyzed during the current study are available from the corresponding author upon reasonable request. Additional raw voltage time-series data and analysis scripts used to identify telegraph switching and metastable attractor states can be provided for verification and reproducibility purposes. The dataset corresponding to Figure~1 is publicly available on Zenodo at \url{https://doi.org/10.5281/zenodo.14716955}.

%\bibliographystyle{unsrt}
%\bibliography{references}

\end{document}